\pdfoutput=1
\documentclass[a4paper,12pt,english]{article}

\usepackage[pdftex]{graphicx}
\usepackage{amsmath}
\usepackage{amssymb}
\usepackage{caption}
\usepackage{appendix}
\usepackage{hyperref} 
\usepackage{multirow}
\usepackage{float}
\usepackage{mathtools}
\usepackage{comment}
\usepackage{subcaption}

\usepackage{xcolor}
\usepackage{hyperref}

\hypersetup{
    colorlinks=true,
    linkcolor=blue,      
    citecolor=blue,     
    urlcolor=blue         
}

\usepackage{wasysym}
\usepackage[style=numeric-comp,sorting=none,maxbibnames=5, minnames=3]{biblatex}
\usepackage{microtype}

\usepackage{color}
\definecolor{gray}{rgb}{0.5,0.5,0.5}

\def\etal{\hbox{\it et al.}}

\newcommand\lsim{\mathrel{\rlap{\lower4pt\hbox{\hskip1pt$\sim$}}
    \raise1pt\hbox{$<$}}}
\newcommand\gsim{\mathrel{\rlap{\lower4pt\hbox{\hskip1pt$\sim$}}
    \raise1pt\hbox{$>$}}}

\newcommand{\beq}{\begin{equation}}
\newcommand{\eeq}{\end{equation}}
\newcommand{\bea}{\begin{eqnarray}}
\newcommand{\eea}{\end{eqnarray}}
\newcommand{\bem}{\begin{pmatrix}}
\newcommand{\eem}{\end{pmatrix}}
\newcommand{\noi}{\noindent}
\newcommand{\non}{\nonumber}

\newcommand{\bet}{\begin{itemize}}
\newcommand{\eet}{\end{itemize}}
\newcommand{\ben}{\begin{enumerate}}
\newcommand{\een}{\end{enumerate}}

\begin{document}

\numberwithin{equation}{section}

\begin{flushright}
\end{flushright}

\bigskip

\begin{center}

{\Large\bf Infinite Heat Order   in  3+1 Dimensions}

\vspace{1cm}

\centerline{Borut Bajc$^{a}$\footnote{borut.bajc@ijs.si}, Giulia Muco$^{b}$\footnote{giulia@qtc.sdu.dk},   
Francesco Sannino$^{b,c}$\footnote{sannino@cp3.sdu.dk}, and
Sophie Wagner$^{b}$\footnote{sowag21@student.sdu.dk}}
\vspace{0.5cm}

\centerline{$^{a}$ {\it\small J.\ Stefan Institute, 1000 Ljubljana, Slovenia}}
\centerline{$^{b}$ {\it \small Quantum Theory Center ($\hslash$QTC) at IMADA \& D-IAS, Southern Denmark University,}} \centerline{\it \small Campusvej 55, 5230 Odense M, Denmark}
\centerline{$^{c}$ {\it \small Dipartimento di Fisica, E. Pancini, Univ. di Napoli, Federico II and INFN sezione di Napoli.}} \centerline{\it \small Complesso Universitario di Monte S. Angelo Edificio 6, via Cintia, 80126 Napoli, Italy}

\end{center}

\bigskip

\begin{abstract}
We investigate whether spontaneous symmetry breaking can persist up to arbitrarily high temperature in ultraviolet-complete quantum field theories in four spacetime dimensions. We focus on completely asymptotically free models with gauge group $\mathrm{SU}(N_{c1})\times \mathrm{SU}(N_{c2})$ and two complex scalar fields, each transforming in the fundamental representation of one gauge factor and singlet under the other. The scalar potential contains quartic self-interactions together with a negative portal coupling between the two sectors. In the Veneziano limit, this class of theories was previously shown to admit fixed-flow trajectories for which one scalar acquires a negative thermal mass at asymptotically large temperature, leading to symmetry non-restoration. Here we extend that analysis to finite numbers of colours and flavours. We derive the finite-$N$ fixed-flow equations, compute the leading $1/N$ corrections to the large-$N$ solutions, and solve the full finite-$N$ system numerically. We find explicit finite-$N$ benchmark theories for which the scalar potential remains bounded from below, the gauge sector is asymptotically free, and one scalar thermal mass stays negative at arbitrarily high temperature. This provides an explicit perturbative example of infinite heat order in a four-dimensional ultraviolet-complete quantum field theory with a finite field content.
\end{abstract}

\clearpage


\section{Introduction}
\label{sec:intro}

Does spontaneous symmetry breaking survive at arbitrarily high temperatures? Standard thermodynamic intuition suggests a negative answer: in the free energy $F=E-TS$, the entropic contribution typically dominates at large $T$, favouring disordered (symmetric) phases. Quantum field theory (QFT) can evade this expectation. Weinberg showed long ago that multi-scalar theories can display high-temperature symmetry breaking through negative thermal masses induced by interactions \cite{Weinberg:1974hy}. This phenomenon is known as \emph{symmetry non-restoration}. A modern perspective was provided by Han, Huang, Komargodski, Lucas and Popov, who identified a general mechanism termed \emph{entropic order}: ordering one sector can increase the available phase space of another sector, so that the ordered phase can carry more entropy than the disordered one \cite{Han:2025eiw}.

The possibility of symmetry non-restoration was developed in particle-physics settings by Mohapatra and Senjanovi\'c \cite{Mohapatra:1979qt,Mohapatra:1979vr}. Cosmological applications include potential resolutions of the monopole \cite{Langacker:1980kd,Salomonson:1984rh,Dvali:1995cj}, domain-wall \cite{Dvali:1995cc} and false-vacuum problems, as well as implications for baryogenesis \cite{Mohapatra:1979zc,Kuzmin:1981bc,Kuzmin:1981ip,Kuzmin:1982hy,Dodelson:1989ii,Dodelson:1991iv} and inflation \cite{Lee:1995fb}. Symmetry non-restoration can also be driven by large conserved charges, through Bose--Einstein condensation or superconducting phases \cite{Linde:1976kh,Haber:1981ts,Benson:1991nj,Riotto:1997tf,Liu:1993am,Bajc:1997ky,Bajc:1998rd,Bajc:1999he}, with phenomenological motivations ranging from neutrino physics to beyond-the-Standard-Model dynamics \cite{Kang:1991xa,Kinney:1999pd,Lesgourgues:1999wu,Barenboim:2016shh,Barenboim:2016lxv,Barenboim:2017dfq}. Supersymmetric theories generically restore their symmetries at high temperature \cite{Haber:1982nb,Mangano:1984dq,Bajc:1996kj}, except in the presence of flat directions \cite{Dvali:1998ct,Bajc:1998jr} or fixed charges. For non-supersymmetric models, the phenomenon has been investigated with a range of perturbative and non-perturbative techniques \cite{Bimonte:1995sc,Orloff:1996yn,Roos:1995vm,Pietroni:1996zj,AmelinoCamelia:1996sd,Pinto:1999pg,Bimonte:1995xs}; some claims remain controversial \cite{Fujimoto:1984hr,Klimenko:1988ng,Grabowski:1990qc,Gavela_1998,Bimonte:1998he}. Extensions to lower and non-integer dimensions were also explored \cite{Hong:2000rk,Chai:2020zgq,Buchel:2020thm,Buchel:2020jfs,Chai:2020onq}; see \cite{Senjanovic:1998xc,Bajc:1999cn} for reviews. The topic was also explored via the AdS/CFT correspondence
\cite{Buchel:2021ead,Buchel:2021yay,Buchel:2022zxl,Buchel:2023zpe} and further discussed within non-local theories in \cite{Chai:2021tpt,Chai:2021djc}. Examples with an infinite number of fields can be found in \cite{Chai:2020hnu,Bajc:2021,Nakayama:2021fgy,Chaudhuri:2021dsq}.
Finally, spontaneous breaking of time-reversal symmetry was investigated in \cite{Carlstr_m_2015}.

Despite this extensive literature, except for large-$N$ examples such as those in \cite{Chai:2020hnu, Bajc:2021},  essentially all four-dimensional examples have been \emph{effective field theories}. Scalar self-interactions (and often Yukawa couplings) grow in the ultraviolet and typically hit a Landau pole, so the theory ceases to exist above some cutoff. In such a situation, statements about the strict $T\to\infty$ limit are not meaningful: the thermal scale inevitably exceeds the regime of validity of the effective description.

A sharp formulation of the problem, therefore, requires UV-complete theories. In Wilson's renormalisation-group picture \cite{Wilson:1971bg,Wilson:1971dh}, a local QFT is well-defined at arbitrarily short distances only if its RG flow approaches a UV fixed point. In four dimensions, the best-understood possibilities are a Gaussian fixed point (asymptotic freedom) \cite{Gross:1973id,Politzer:1973fx} or an interacting fixed point (asymptotic safety) \cite{Weinberg:1976xy}. Only in such theories can one consistently send the temperature to arbitrarily large values without leaving the domain of the microscopic theory.

In our previous work \cite{Bajc:2021} we initiated a systematic study of infinite heat order in UV-complete four-dimensional gauge-Yukawa theories. Working mainly in the Veneziano limit, we found that complete asymptotic freedom can coexist with symmetry non-restoration in models with two gauged scalars transforming under a product gauge group $\mathrm{SU}(N_{c1})\times \mathrm{SU}(N_{c2})$, while the asymptotically safe examples studied there displayed high-temperature symmetry restoration.

Related progress was obtained recently in lower dimensions. Komargodski and Popov \cite{Komargodski:2024zmt} constructed a UV-complete, local and unitary QFT in $2+1$ dimensions with a finite field content that breaks a $\mathbb{Z}_2$ symmetry at any temperature. Shortly thereafter, Han \etal\ \cite{Han:2025eiw} demonstrated that the underlying mechanism, entropic order, arises broadly across lattice systems and continuum QFT and provides a unifying explanation for high-temperature ordering phenomena.

In this paper, we provide an explicit UV-complete example in $3+1$ dimensions with a finite number of degrees of freedom that exhibits spontaneous symmetry breaking at arbitrarily high temperature. The theory is completely asymptotically free: all couplings flow to a Gaussian UV fixed point, so the microscopic description remains valid at all scales. The analysis is perturbative and controlled by one-loop RG evolution together with the high-temperature effective potential.

Our minimal model contains two complex scalars $\varphi_1$ and $\varphi_2$, each transforming in the fundamental representation of an independent non-abelian gauge group $\mathrm{SU}(N_{c1})$ and $\mathrm{SU}(N_{c2})$, respectively. The renormalisable scalar potential includes a quartic cross-coupling $\lambda(\varphi_1^\dagger\varphi_1)(\varphi_2^\dagger\varphi_2)$ analogous to Weinberg's original setup \cite{Weinberg:1974hy}.
Imposing the fixed-flow conditions required for complete asymptotic freedom restricts the couplings to a surface parametrised by the ratio $x=N_{c1}/N_{c2}$. For an appropriate range of $x$, one of the scalar thermal masses becomes negative asymptotically, leading to symmetry non-restoration. We demonstrate this mechanism both in the Veneziano limit and at finite $N_{ci}$, combining analytic $1/N_{ci}$ corrections with a direct numerical solution of the finite-$N$ RG equations. We also exhibit an explicit anomaly-free benchmark point at finite $N$, confirming that the construction involves a genuinely finite field content.

The product gauge-group structure is essential. Models with a single $\mathrm{SU}(N_c)$ gauge factor and fundamental scalars, as well as the singlet-scalar class studied in \cite{Bajc:2021}, restore their symmetries at high temperature once complete asymptotic freedom is imposed. Two independent gauge sectors provide sufficient parametric freedom for the fixed-flow constraints and the negative-thermal-mass condition to be simultaneously satisfied, thereby realising the entropic-order mechanism of \cite{Han:2025eiw} in four dimensions.

The paper is organised as follows. Section \ref{sec:LOresults} reviews the large-$N$ analysis of \cite{Bajc:2021}, including Weinberg's two-scalar model and the $\mathrm{SU}(N_{c1})\times\mathrm{SU}(N_{c2})$ model exhibiting infinite heat order in the Veneziano limit. Section \ref{sec:finiteN} extends the analysis to finite $N_{ci}$, providing analytic next-to-leading corrections and numerical solutions. Section \ref{sec:conclusions} contains our conclusions. The appendices collect one-loop RG equations and thermal-mass formulae for generic gauge-Yukawa theories, together with details for the specific models analysed in the main text.

\section{Infinite Heat Order with Finite Degrees of Freedom }
\subsection{The Infinite Degrees of Freedom Limit Recap}
\label{sec:LOresults}
 We start by briefly reviewing the results in \cite{Bajc:2021},  where symmetry non-restoration at infinite temperature (infinite heat order) was first discovered in four dimensions within asymptotically free theories. It generalises the results of Weinberg  \cite{Weinberg:1974hy}, which is valid for effective theories, meaning theories with an ultraviolet cutoff. Weinberg's original example \cite{Weinberg:1974hy} features two scalars interacting via a quartic potential with a cross-coupling term that preserves a discrete $\mathbb{Z}_2\times\mathbb{Z}_2$ symmetry:
\begin{equation}
    V=\frac{\lambda_1}{4}\phi^4_1+\frac{\lambda_2}{4}\phi_2^4-\frac{\lambda}{2}\phi_1^2\phi_2^2\;. \label{eq:VfromWeinberg}
\end{equation}
For the model to be viable, the potential has to be bounded from below, imposing constraints on the couplings
\begin{align}
    \lambda_{1,2}>0\quad,\quad \lambda^2<\lambda_1\lambda_2\,.\label{eq:Vbounded}
\end{align}
At high temperatures, the potential acquires a leading-order thermal correction
\begin{equation}
    \Delta V_T= \frac{T^2}{24}((3\lambda_1-\lambda)\phi_1^2 +(3\lambda_2-\lambda)\phi^2_2)\;,
\end{equation}
which for $\lambda>3\lambda_2$ will ensure that $\phi_2$ acquires a negative thermal mass, resulting in a non-zero VEV at high temperatures. However, as mentioned above, this model is not UV complete, since the required large cross-coupling increases with the energy scale, eventually reaching a Landau pole. 

In this work, we close the gap and show that we can extend asymptotically free models supporting infinite heat order \cite{Bajc:2021} to a finite number of degrees of freedom. The model introduced in \cite{Bajc:2021} consists of the gauge-group product $\mathrm{SU}(N_{c1})\times\mathrm{SU}(N_{c2})$,  featuring two complex scalar fields $\varphi_i$ each transforming according to the fundamental representation of their respective gauge group $\mathrm{SU}(N_{ci})$ while staying a singlet under the other. The most general renormalizable potential is
\begin{align}
    V=\frac{\lambda_1}{2}\left(\vec{\varphi}_1^*\cdot\vec{\varphi}_1\right)^2+\frac{\lambda_2}{2}\left(\vec{\varphi}_2^*\cdot\vec{\varphi}_2\right)^2-
\lambda\left(\vec{\varphi}_1^*\cdot\vec{\varphi}_1\right)\left(\vec{\varphi}_2^*\cdot\vec{\varphi}_2\right)\;,
\end{align}
which naturally mimics that of Weinberg's model \eqref{eq:VfromWeinberg} by including the cross-coupling term between the two sectors, which is essential for generating a negative mass squared. The global symmetry over the scalars is still the Weinberg one.

To verify the presence of symmetry non-restoration, we first consider the effective potential $\Delta V_T$ at large $N_{ci}$. The logic is simple: we impose fixed-flow trajectories, solve the resulting algebraic constraints for the rescaled couplings, and then inspect the sign of the thermal masses. If one of them is negative while the zero-temperature potential remains bounded from below, then the corresponding scalar condenses even at asymptotically high temperature.

In this process, we parametrise the couplings in terms of RG time $t=\log(\mu/\mu_0)$ to enforce the fixed flow solution. Further, the couplings are scaled by $N_{ci}$, to ensure stable RGEs in the Veneziano limit, and are given by
\begin{align}
    g_i^2=\frac{16\pi^2\tilde\alpha_i}{N_{ci}t}\quad,\quad \lambda_i=\frac{16\pi^2\tilde\lambda_i}{N_{ci}t}\quad,\quad\lambda=\frac{16\pi^2\tilde\lambda}{\sqrt{N_{c1}N_{c2}}t}\;,
\end{align}
for $i=1,2$. We further introduce the variables $\tilde\lambda_\pm=\frac{1}{2}\left(\tilde\lambda_1\pm\tilde\lambda_2\right)$, and $\tilde\alpha_\pm=\frac{1}{2}\left(\tilde\alpha_1\pm\tilde\alpha_2\right)$. Using these variables and enforcing fixed flow trajectories, the RGEs yield the following solutions
\beq\label{eq:LO_sol}
    \tilde\alpha_-=0 \quad,\quad
\tilde\lambda_+=\frac{6\tilde\alpha_+-1}{4}\quad,\quad
\tilde\lambda_-^2+\tilde\lambda^2=\frac{1}{16}\left(24\tilde\alpha_+^2-12\tilde\alpha_++1\right)\;,
\eeq
which are valid for
\begin{align}
    \tilde\alpha_+\geq(3+\sqrt{3})/12\;.
\end{align}
Using positive $\tilde\lambda=|\tilde\lambda|$ and considering the negative root for $\tilde\lambda_-$
\begin{equation}
    \label{eq:lamndaminneg}
    \tilde\lambda_-=-\sqrt{\frac{24\tilde\alpha_+^2-12\tilde\alpha_++1}{16}-|\tilde\lambda|} \ \  ,
\end{equation}
from the effective potential $\Delta V_T$ the thermal mass parameters $\mu_1^2$ and $\mu_2^2$ are extracted 
\begin{align}
    \mu_1^2&=\frac{12\tilde\alpha_+-1}{2}-2\left(\sqrt{\frac{24\tilde\alpha_+^2-12\tilde\alpha_++1}{16}-|\tilde\lambda|^2}+\sqrt{\frac{N_{c2}}{N_{c1}}}|\tilde\lambda|\right)\;,
    \\
    \mu_2^2&=\frac{12\tilde\alpha_+-1}{2}+2\left(\sqrt{\frac{24\tilde\alpha_+^2-12\tilde\alpha_++1}{16}-|\tilde\lambda|^2}-\sqrt{\frac{N_{c1}}{N_{c2}}}|\tilde\lambda|\right)\;.
\end{align}
Here, the last term is the crucial one: the negative contribution proportional to $\sqrt{N_{c1}/N_{c2}}\,|\tilde\lambda|$ can dominate for sufficiently small colour ratio $x=N_{c1}/N_{c2}$. The large-$N$ analysis, therefore, already isolates the basic mechanism that will survive at finite $N$: the portal interaction can overcome the positive gauge and quartic contributions in one scalar channel while keeping the potential bounded from below.

Checking the second condition \eqref{eq:Vbounded} for a potential bounded from below $\lambda_1\lambda_2-\lambda^2>0$ and ensuring real fixed flow solutions, we find a constraint on the ratio of the number of fermions to colours
\begin{align}
    \frac{1}{2}\left(2+3\sqrt{3}\right)\leq\frac{N_{fi}}{N_{ci}}<\frac{11}{2}\;.
\end{align}
Ultimately, this framework successfully yields a UV-complete model where one thermal mass turns negative for a given ratio $x$. Stability of the vacuum and the requirement for real algebraic solutions to the RGEs necessitate a specific bound on the ratio of fermions to colours, alongside a symmetric gauge structure $\tilde\alpha_1=\tilde \alpha_2$. This preserves the broken symmetry at high energies\footnote{Note that this model also has an IR Banks-Zaks fixed circle. Previous analysis \cite{Bajc:2021} found that symmetry non-restoration could also be realised near this IR fixed point. }. 

The analytical success of the model with two scalars, respectively gauged under $\mathrm{SU}(N_{c1})\times\mathrm{SU}(N_{c2})$, was achieved by simplifications of the Veneziano limit. This limit decoupled and simplified the RGEs. A relevant question to ask is whether the characteristic of negative thermal mass persists beyond this limit. Thus, in the next section, we will identify the parameter space for a negative thermal mass in the case of finite $N$. 

\subsection{Finite number of degrees of freedom}\label{sec:finiteN}

Having established the general conditions under which infinite heat order can arise, we now turn to the more restrictive and physically relevant case in which the number of degrees of freedom is finite. This setting is important for two reasons. First, it shows that the phenomenon is not tied to large-$N$ idealisations or to parametrically large field multiplicities. Second, it allows one to identify explicitly the minimal structures required for the persistence of ordered phases at arbitrarily high temperatures. The analysis in this section, therefore, serves to clarify how much of the mechanism survives once the theory is reduced to a genuinely finite field content.

Following the derivation in Appendix \ref{SUNc12} and introducing the parameter $x$ through $N_{c1} = x\,N_{c2}$, the finite-$N$ RG equations take the form
\begin{align}\label{eq:finiteNRG}
-\tilde{\lambda}_1 \;=&\;
\left(2+\frac{8}{x N_{c2}}\right)\tilde{\lambda}_1^{\,2}
+2\,\tilde{\lambda}^{\,2}
-6\!\left(1-\frac{1}{x^{2}N_{c2}^{2}}\right)\tilde{\alpha}_1\tilde{\lambda}_1
\nonumber\\&+\frac{3}{2}\tilde{\alpha}_1^{\,2}\!\left(
1+\frac{1}{x N_{c2}}-\frac{4}{x^{2}N_{c2}^{2}}+\frac{2}{x^{3}N_{c2}^{3}}
\right),\nonumber
\\[4pt]
-\tilde{\lambda}_2 \;=&\;
\left(2+\frac{8}{N_{c2}}\right)\tilde{\lambda}_2^{\,2}
+2\,\tilde{\lambda}^{\,2}
-6\!\left(1-\frac{1}{N_{c2}^{2}}\right)\tilde{\alpha}_2\tilde{\lambda}_2
\nonumber\\&+\frac{3}{2}\tilde{\alpha}_2^{\,2}\!\left(
1+\frac{1}{N_{c2}}-\frac{4}{N_{c2}^{2}}+\frac{2}{N_{c2}^{3}}
\right),
\\[4pt]
-\tilde{\lambda} \;=&\;
\tilde{\lambda}\Bigg[
-\frac{4}{N_{c2}\sqrt{x}}\,\tilde{\lambda}
+2\!\left(1+\frac{1}{x N_{c2}}\right)\tilde{\lambda}_1
+2\!\left(1+\frac{1}{N_{c2}}\right)\tilde{\lambda}_2
\nonumber\\&-3\!\left(
\tilde{\alpha}_1\!\left(1-\frac{1}{x^{2}N_{c2}^{2}}\right)
+\tilde{\alpha}_2\!\left(1-\frac{1}{N_{c2}^{2}}\right)
\right)
\Bigg] \ . \nonumber
\end{align}
Moreover, the finite-$N$ thermal potential is
\begin{equation}\label{eq:potential}
    \begin{split}
        \Delta V_T&=
(4\pi)^2\frac{T^2}{24\log{T}}\left\{\left[2\left(\tilde\lambda_1\left(1+\frac{1}{N_{c1}}\right)-\sqrt{\frac{N_{c2}}{N_{c1}}}\tilde\lambda\right)+3\tilde\alpha_1 \left(1 - \frac{1}{N_{c1}^2} \right)\right]
\left(\vec{\varphi}_1^*\cdot\vec{\varphi}_1\right)\right.\\
&\left.+\left[2\left(\tilde\lambda_2\left(1+\frac{1}{N_{c2}}\right)-\sqrt{\frac{N_{c1}}{N_{c2}}}\tilde\lambda\right)+3\tilde\alpha_2 \left(1 - \frac{1}{N_{c2}^2} \right)\right]
\left(\vec{\varphi}_2^*\cdot\vec{\varphi}_2\right)\right\} \ ,
    \end{split}
\end{equation}
yielding quadratic thermal masses
\begin{equation}
    \label{eq:finiteNmass}
     \mu^2_i= 3 \tilde\alpha_i  \left(1-\frac{1}{N_{ci}^2}\right)+2 \tilde\lambda_i  \left(\frac{1}{N_{ci}}+1\right) -\, 2\sqrt{\frac{N_{cj}}{N_{ci}}}\,\tilde\lambda \ ,
\end{equation}
for $(i,j)=(1,2), \ (2,1)$.\\
These expressions make the finite-$N$ problem concrete. Once the algebraic RG constraints are solved, the sign of $\mu_i^2$ can be read off directly from \eqref{eq:finiteNmass}. The strategy in the remainder of this section is therefore to proceed in two complementary ways: first by constructing the leading $1/N_{ci}$ deformation of the Veneziano-limit solution, and then by solving the full finite-$N$ equations numerically.\\

\subsubsection{Symmetry non-restoration at NLO}
\begin{figure}[h!]
\begin{center}
\includegraphics[width=11cm]{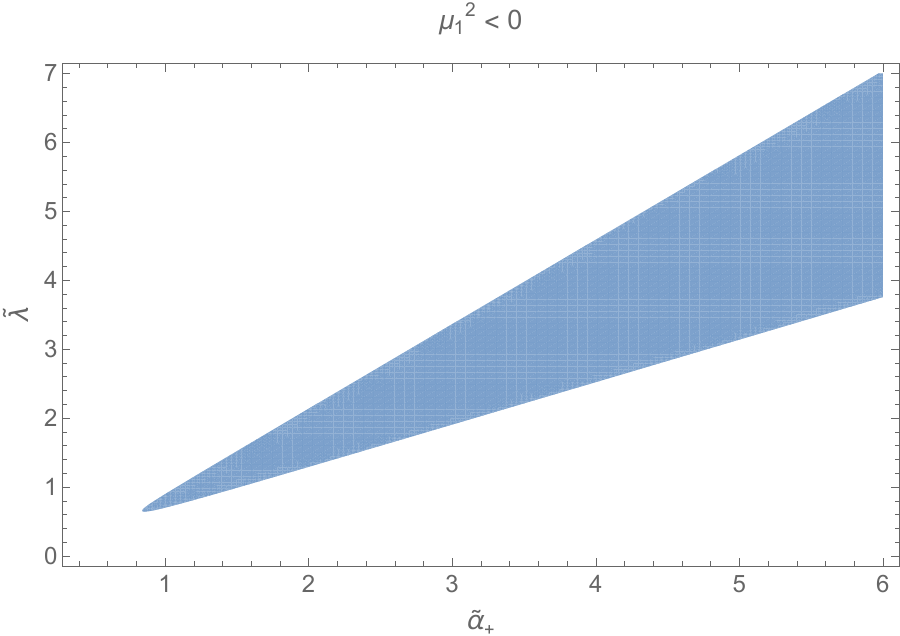}
\end{center}
\caption{\label{fig:regionplot} The shadowed region shows where in the $\tilde\alpha_+-\tilde\lambda$ parameter space the squared mass of $\varphi_1$ is negative. Here we set $\tilde{\alpha}_+^{(1)}=\tilde{\lambda}^{(1)}=0$ (as allowed by the RG equations), $x=0.1$ and $N_{c2}=1000$.}
\end{figure}

  The goal of this subsection is to exhibit analytically how the relevant thermal contributions reorganise themselves in a finite system, and to determine under which conditions the ordered phase can be maintained as the temperature increases. This provides the simplest concrete realisation of infinite heat order at finite $N$, and it will serve as a useful benchmark for the more general constructions discussed below. Our roadmap is the following: we expand the couplings around the Veneziano-limit solution, solve the $1/N_{ci}$ algebraic constraints, and finally insert the result into the thermal mass in order to determine whether the ordered phase persists.
In the finite $N_{ci}$ system of equations \eqref{eq:finiteNRG}, we redefine the couplings defined in Section \ref{sec:LOresults} by separating their LO and NLO terms, i.e. 
\begin{equation}\label{eq:couplings}
\begin{split}
    \tilde\lambda_\pm&=   \tilde\lambda^{(0)}_\pm+\frac{1}{N_{c2}}   \tilde\lambda_\pm^{(1)} \ ,\\
        \tilde \lambda&=   \tilde\lambda^{(0)}+\frac{1}{N_{c2}}   \tilde\lambda^{(1)} \ ,\\
          \tilde\alpha_\pm&=   \tilde\alpha^{(0)}_\pm+\frac{1}{N_{c2}}   \tilde\alpha_\pm^{(1)} \ .\\
\end{split}
\end{equation}
Here the colour ratio $x=N_{c1}/N_{c2}$ is kept fixed, while $1/N_{c2}$ is the actual expansion parameter. By plugging these expressions into the finite-$N$ RG equations and truncating at order $1/N_{c2}$, the coefficients $\tilde\lambda^{(0)}, \ \tilde\lambda_\pm^{(0)}, \ \tilde\alpha^{(0)}_\pm$ are identified with the large-$N$ solutions \eqref{eq:LO_sol}-\eqref{eq:lamndaminneg} of the LO equations \eqref{eq:LORGeq1}-\eqref{eq:LORGeq3}. The $1/N_{c2}$ terms then determine the NLO corrections $\tilde\lambda^{(1)}, \ \tilde\lambda_\pm^{(1)}, \ \tilde\alpha^{(1)}_\pm$, shown in equations~\eqref{eq:NLORGeqp}-\eqref{eq:NLORGeq3}. From these, we find 
\begin{align}\label{eq:am1Sol}
        \tilde\alpha_-^{(1)}&=\frac{1}{12x(4\tilde\alpha_+^{(0)}-1)}\Big[3-16\tilde\lambda^{(0)2}+3K-8\sqrt{x}\tilde\lambda^{(0)}K+6\tilde\alpha_+^{(0)}\Big(17\tilde\alpha_+^{(0)}-3(2K)\Big)\\\nonumber &+x\Big(-3+16\tilde\lambda^{(0)2}+3K-6\tilde\alpha_+^{(0)}(-6+17\tilde\alpha_+^{(1)}+3K)\Big)\Big] \ \ \ ,\\
        \label{eq:l1pSol}
    \tilde\lambda_+^{(1)}&=\frac{1}{8x}\Big[1+x-6(1+x)\tilde\alpha_+^{(0)}+12x\tilde\alpha_+^{(1)} +8\sqrt{x}\tilde\lambda_+^{(0)}+(1-x)K\Big] \ \ \ , \\\label{eq:lm1Sol}
    \tilde\lambda_-^{(1)}&=\frac{1}{4xK}\Big[63\tilde\alpha_+^{(0)2}-12\tilde\alpha_+^{(0)}(2+K)+2(1-8\tilde\lambda^{(0)2}+K)+x(2\\\nonumber
    &+6\tilde\alpha_+^{(1)}-2K+6\tilde\alpha_-^{(1)}K+3\tilde\alpha_+^{(0)}(-8+21\tilde\alpha_+^{(0)}-8\tilde\alpha_+^{(1)}+4K)\\\nonumber
    &+16\tilde\lambda^{(0)}(\tilde\lambda^{(1)}-\tilde\lambda^{(0)}))\Big] \ \ \ ,
\end{align}
where $K=\sqrt{1+12\tilde\alpha_+^{(0)}(2\tilde\alpha_+^{(0)}-1)-16\tilde\lambda^{(0)2}}$.\\
The final step to see that symmetry non-restoration persists with a finite number of fields is to plug the LO and NLO solutions into the thermal mass \eqref{eq:finiteNmass} and truncate at order $1/N_{c2}$. As an example, we consider the thermal mass \(\mu_1^2\). Without loss of generality, we set $\tilde\alpha_+^{(1)}=\tilde\lambda^{(1)}=0$, as allowed by the RG equations, and therefore drop the superscripts on $\tilde\alpha_+^{(0)}$ and $\tilde\lambda^{(0)}$. This gives
\begin{equation}
\label{eq:mT2FiniteN} 
\begin{split} 
\mu^2_1&=3\tilde\alpha_+-\frac{2\tilde\lambda}{\sqrt{x}}+\frac{1}{2x^2N_{c2}^2(4\tilde\alpha_+-1)K}\Bigg\{(1+x N_{c2})\Big[-1 - 12 ( x-7) \tilde\alpha_+^3 - K\\
&- x N_{c2} ( 4 \tilde\alpha_+-1) \Big(1 + 24 \tilde\alpha_+^2 - 16 \tilde\lambda^2 + K - 6 \tilde\alpha_+ (2 + K)\Big) + 4 \tilde\lambda \Big(-2 \tilde\lambda K\\ 
&+ 2 x \tilde\lambda ( K-2) - \sqrt{x} (1 - 16 \tilde\lambda^2 + K)\Big) + \tilde\alpha_+ \Big(13 + 48 \tilde\lambda^2 + 7 K + 16 \sqrt{x} \tilde\lambda ( K+3) \\
&+ x (-3 + 112 \tilde\lambda^2 + 3 K)\Big) - 3 \tilde\alpha_+^2 \Big(17 + 32 \sqrt{x} \tilde\lambda + 3 K + x (-7 + 5 K)\Big)\Big]\Bigg\} \ \ . 
\end{split} 
\end{equation}
In Fig.\ \ref{fig:regionplot}, we show that there exists a region in the $\tilde\alpha_+-\tilde\lambda$ parameter space where this thermal mass becomes negative.

\subsubsection{On the lower bound on the colour degrees of freedom}

Having established that symmetry non-restoration survives at subleading order in the inverse number of colours, we now determine, within perturbation theory, the lowest possible bound on $N_{c1}$ and $N_{c2}$ below which symmetry must restore at high temperature. To analyse this situation, we recall the equations guaranteeing UV freedom in Eq.\ \eqref{eq:finiteNRG} at a finite number of colours as well as that the leading high temperature  1-loop correction to the potential is \eqref{eq:potential}, with the dimensionless thermal masses given in \eqref{eq:finiteNmass}.
The symmetry non-restoration condition implies  

\beq
\mu_1^2\equiv-m^2\leq0
\eeq
\noi
where, to ease notation, we introduced the positive $m^2$ parameter and start computing $\tilde\lambda$ from $\mu_1^2$ in Eq.\ \eqref{eq:finiteNmass}. This yields:
\beq
\tilde\lambda=\sqrt{\frac{N_{c1}}{N_{c2}}}\left(\frac{m^2}{2}+\tilde\lambda_1\left(1+\frac{1}{N_{c1}}\right)+\frac{3}{2}\tilde\alpha_1\left(1-\frac{1}{N_{c1}^2}\right)\right) \ .
\eeq

\noi
We then substitute it into the first RG equation in the system (\ref{eq:finiteNRG}). It is  useful to transform the variables $\tilde\lambda_1$, $\tilde\alpha_1$ into $\ell_1$, $a_1$ via 

\beq
\tilde\lambda_1=\ell_1-a(N_{c1},N_{c2}) a_1-b(N_{c1},N_{c2},m^2)\quad,\quad \tilde\alpha_1=a_1+c(N_{c1},N_{c2},m^2) \ ,
\eeq
with  $a$, $b$, $c$ functions constructed in the following way, such that we can eliminate $\ell_1a_1$, $\ell_1$ and $a_1$ terms in the first RG equation of \eqref{eq:finiteNRG}:
\begin{equation}
    a(N_{c1},N_{c2})=\frac{3 \left(N_{c1}^2-1\right) \left(N_{c1}-N_{c2}+1\right)}{2 N_{c1} \left(\left(N_{c1}+1\right){}^2+\left(N_{c1}+4\right) N_{c2}\right)} \ ,
\end{equation}
\begin{equation}
    \begin{split}
        &b(N_{c1},N_{c2},m^2)=\\
        &=\frac{N_{c1} \left(N_{c1}+1\right) \left(2 m^2 \left(4 N_{c1}^2+2 N_{c1}-5\right)+3 N_{c1}^2-3\right)+N_{c1} \left(N_{c1} \left(N_{c1}+2\right)-2\right) N_{c2}}{4 N_{c1} \left(9 N_{c2}+N_{c1} \left(N_{c1} \left(10 N_{c1}-2 N_{c2}+31\right)+3 \left(N_{c2}+4\right)\right)-29\right)-20 \left(N_{c2}+4\right)} \ ,
    \end{split}
\end{equation}
\begin{equation}
    \begin{split}
        &c(N_{c1},N_{c2},m^2)=\\
        &=-\frac{N_{c1}^2 \left(N_{c1}+1\right) \left(2 m^2 \left(2 N_{c1}+5\right)-N_{c1}+N_{c2}-1\right)}{2 N_{c1} \left(9 N_{c2}+N_{c1} \left(N_{c1} \left(10 N_{c1}-2 N_{c2}+31\right)+3 \left(N_{c2}+4\right)\right)-29\right)-10 \left(N_{c2}+4\right)} \ .
    \end{split}
\end{equation}
Therefore the first RG equation in the system (\ref{eq:finiteNRG}) now reads
\beq
c_{\ell_1}(N_{c1},N_{c2})\ell_1^2+c_{a_1}(N_{c1},N_{c2})a_1^2=c_1(N_{c1},N_{c2},m^2) \ ,
\label{quadratic}
\eeq

\noi
with
\begin{equation}
    c_{\ell_1}(N_{c1},N_{c2})=\frac{\left(N_{c1}+1\right){}^2+\left(N_{c1}+4\right) N_{c2}}{N_{c1} N_{c2}} \ ,
\end{equation}
\begin{equation}\begin{split}
    &c_{a_1}(N_{c1},N_{c2})=\\
    &=\frac{3 \left(N_{c1}-1\right) \left(\left(N_{c1} \left(10 N_{c1}+11\right)-20\right) \left(N_{c1}+1\right){}^2
+\left(N_{c1} \left(\left(3-2 N_{c1}\right) N_{c1}+9\right)-5\right) N_{c2}\right)}{4 N_{c1}^3 \left(\left(N_{c1}+1\right){}^2+\left(N_{c1}+4\right) N_{c2}\right)} \ ,
\end{split}
\end{equation}
\begin{align}
c_1(N_{c1},N_{c2},m^2)&=\frac{c_1^u(N_{c1},N_{c2},m^2)}{c_1^d(N_{c1},N_{c2})} \ ,\\
c_1^u(N_{c1},N_{c2},m^2)&=N_{c1} \bigg(4 m^4 \left(2 N_{c1}-1\right) \left(\left(N_{c1}-1\right)  
N_{c1}-5\right)\non\\&+4 m^2 \left(N_{c1}+1\right) \left(4 N_{c1}^2+2 N_{c1}-5\right)\bigg) \non\\&
\left.+3 
\left(N_{c1}-1\right) \left(N_{c1}+1\right){}^2+\left(N_{c1} \left(N_{c1}+2
\right)-2\right) N_{c2}\right) \ ,\\
c_1^d(N_{c1},N_{c2})=&16 N_{c1} \left(9 N_{c2}+N_{c1} \left(N_{c1} \left(10 
N_{c1}-2 N_{c2}+31\right)+3 \left(N_{c2}+4\right)\right)-29\right)\nonumber\\&-80 
\left(N_{c2}+4\right)\ .
\end{align}
The solution of the quadratic equation in \eqref{quadratic} can be written as
\bea
\ell_1&=&\sqrt{\frac{c_1}{c_{\ell_1}}}\cos{\phi}\ ,\\
a_1&=&\sqrt{\frac{c_1}{c_{a_1}}}\sin{\phi} \ .
\eea
Additionally, the solution should further satisfy the constraints
\begin{align}
\label{lambda1pozitivna}
\tilde\lambda_1>0&\to\sqrt{\frac{c_1}{c_{\ell_1}}}\cos{\phi}-a\sqrt{\frac{c_1}{c_{a_1}}}\sin{\phi}-b>0\ ,\\
\label{alfa1pozitivna}
\tilde\alpha_1>0&\to\sqrt{\frac{c_1}{c_{a_1}}}\sin{\phi}+c>0\ .
\end{align}
We have that \eqref{lambda1pozitivna} can be satisfied if
\begin{equation}
    \begin{split}
        &c_1\left(\frac{1}{c_{\ell_1}}+\frac{a^2}{c_{a_1}}\right)-b^2=\\
&=-\frac{m^4 N_{c1}^2 \left(N_{c1}^2+2 N_{c1}-2\right)}{4 \left(10 N_{c1}^4+\left(31-2 N_{c2}\right) N_{c1}^3+3 \left(N_{c2}+4\right) N_{c1}^2+\left(9 N_{c2}-29\right) N_{c1}-5 \left(N_{c2}+4\right)\right)}
\geq0 \ .
    \end{split}
\end{equation}

\noi  
This can be satisfied when
\beq
\label{N2}
N_{c2}\geq\frac{10 N_{c1}^4+31 N_{c1}^3+12 N_{c1}^2-29 N_{c1}-20}{2 N_{c1}^3-3N_{c1}^2-9 N_{c1}+5} \ .
\eeq

\begin{figure}[h!]
    \centering
    \begin{subfigure}[b]{0.49\textwidth}
        \centering
        \includegraphics[width=\textwidth]{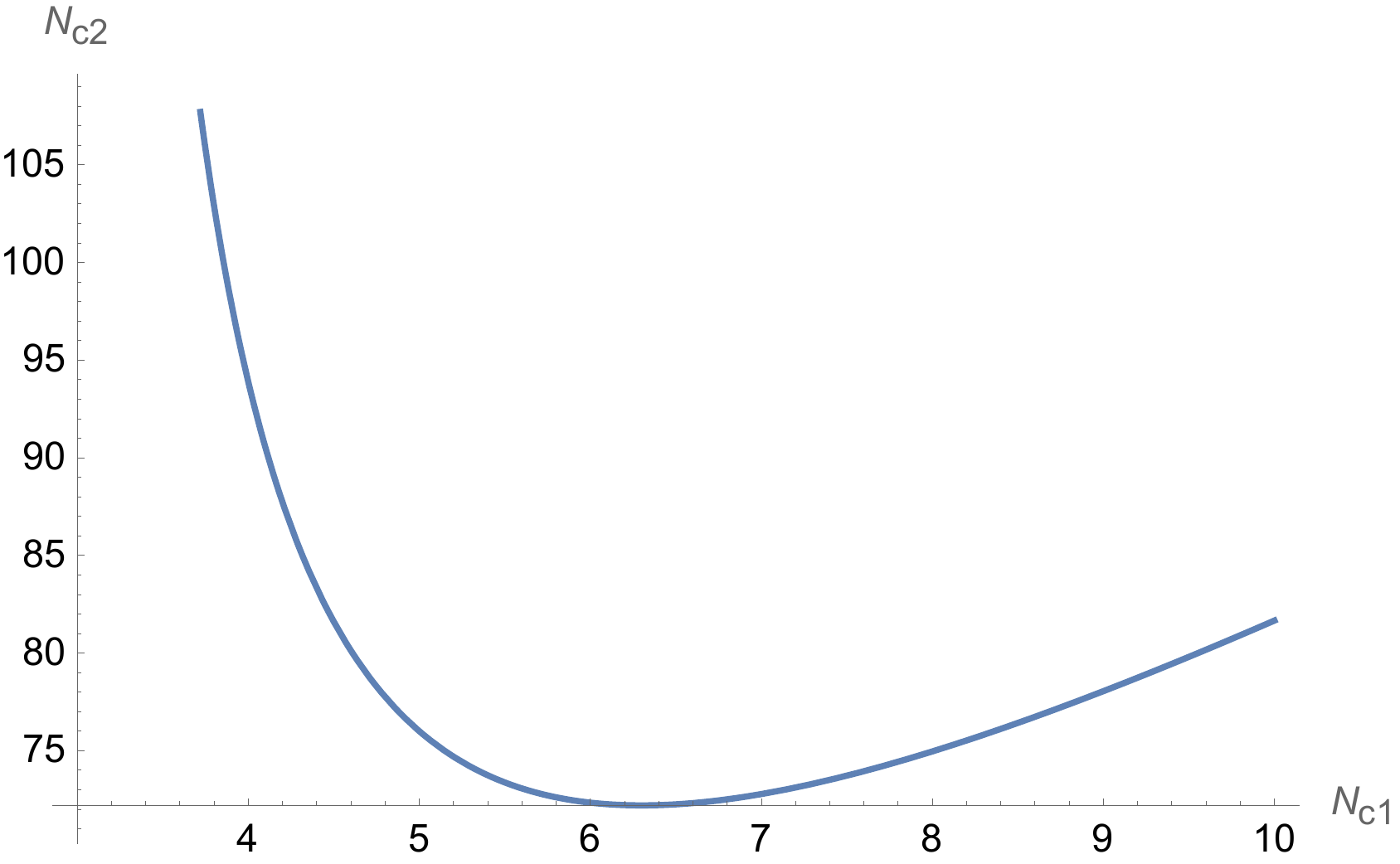}
        \caption{} 
        \label{xvsN1}
    \end{subfigure}
    \hspace{0.05cm}
    \begin{subfigure}[b]{0.49\textwidth}
        \centering
        \includegraphics[width=\textwidth]{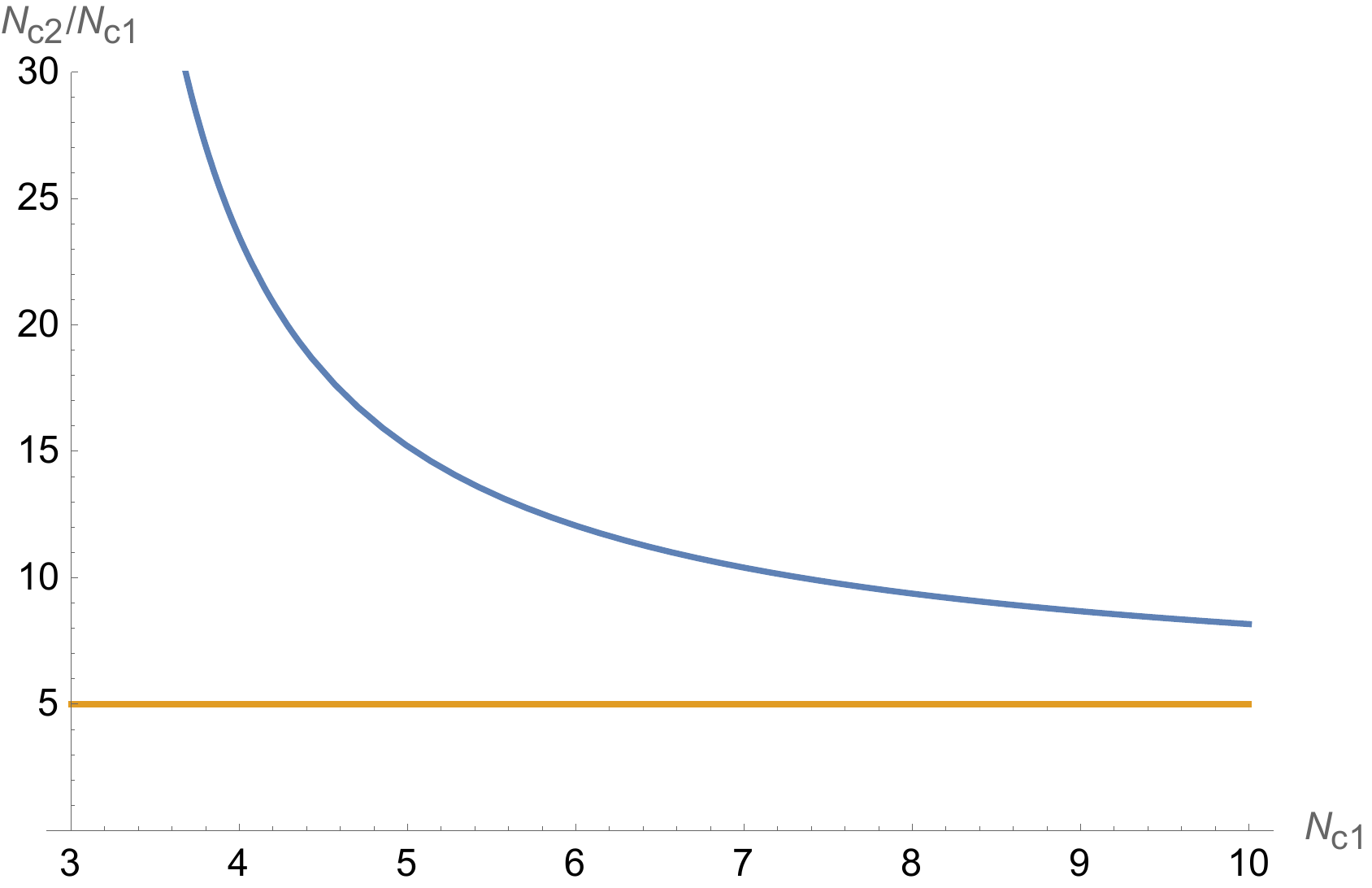}
        \caption{} 
        \label{N2vsN1}
    \end{subfigure}
    
    \caption{(a) The necessary condition for  symmetry non-restoration  to happen in the theory is  that the ratio 
    $N_{c2}/N_{c1}$ lies above the blue curve. The yellow line denotes the lowest possible bound
  for $N_{c2}/N_{c1}$ occurring for $N_{c1}\to\infty$. (b) The necessary condition for the solution to exist is that $N_{c2}$ is above the blue curve.}
    \label{fig:xbound}
\end{figure}

\noi In this range we have $a<0$ and $c>0$, so that  there is a simultaneous solution for   (\ref{lambda1pozitivna})
and (\ref{alfa1pozitivna}) when $\phi$ lies in the first quadrant. We observe that \eqref{N2} implies that $N_{c2} \geq 5 N_{c1}$  with the equality recovered in the limiting case of $N_{c1}\to\infty$ as shown in
 Fig.\ \ref{fig:xbound}(a).

In Fig.~\ref{fig:xbound}(b) we show the lower bound stemming from \eqref{N2} from which it is clear that the lowest possible integer value for $N_{c2}$ is $(N_{c2})_{min}=73$ for either $N_{c1}=6$ or $N_{c1}=7$.

We can therefore conclude that any colour ratio with $x<5$ cannot support symmetry non-restoration. Among the possible pairs of $N_{c1}$ and $N_{c2}$ satisfying \eqref{N2} only those that further abide  
\beq
\tilde\lambda_2>0\quad,\quad\tilde\alpha_2>0\quad,\quad\tilde\lambda_1\tilde\lambda_2>\tilde\lambda^2\  ,
\eeq
will be physical solutions. We will now move to the numerical evaluation of the constraints to identify which physical theory supports symmetry non-restoration.  


\subsubsection{Numerical results}
\label{sec:numerical}

We next extend the analysis numerically to achieve a wider class of finite degrees of freedom theories, within the same class of models, in order to assess the robustness of the mechanism. In particular, this subsection is meant to show which features survive at a smaller number of colours and flavours. This comparison is essential for bridging the gap between analytic results and possible phenomenological future applications.  

We start our analysis here by defining
\beq\label{eq:alpha12def}
\tilde\alpha_{1,2}=\frac{1}{2b_{1,2}}\quad,\quad
b_{1,2}=\frac{11}{3}-\frac{1}{6N_{c1,2}}-\frac{2N_{f1,2}}{3N_{c1,2}}\ .
\eeq

We look for solutions of the finite $N$ RGEs providing

\beq
\tilde\lambda_{1,2}>0\quad,\quad |\tilde\alpha_-|\leq\tilde\alpha_+\quad,\quad \tilde\lambda_1\tilde\lambda_2>\tilde\lambda^2\ .
\eeq

\begin{figure}[h!]
    \centering
    \begin{subfigure}[b]{0.49\textwidth}
        \centering
        \includegraphics[width=\textwidth]{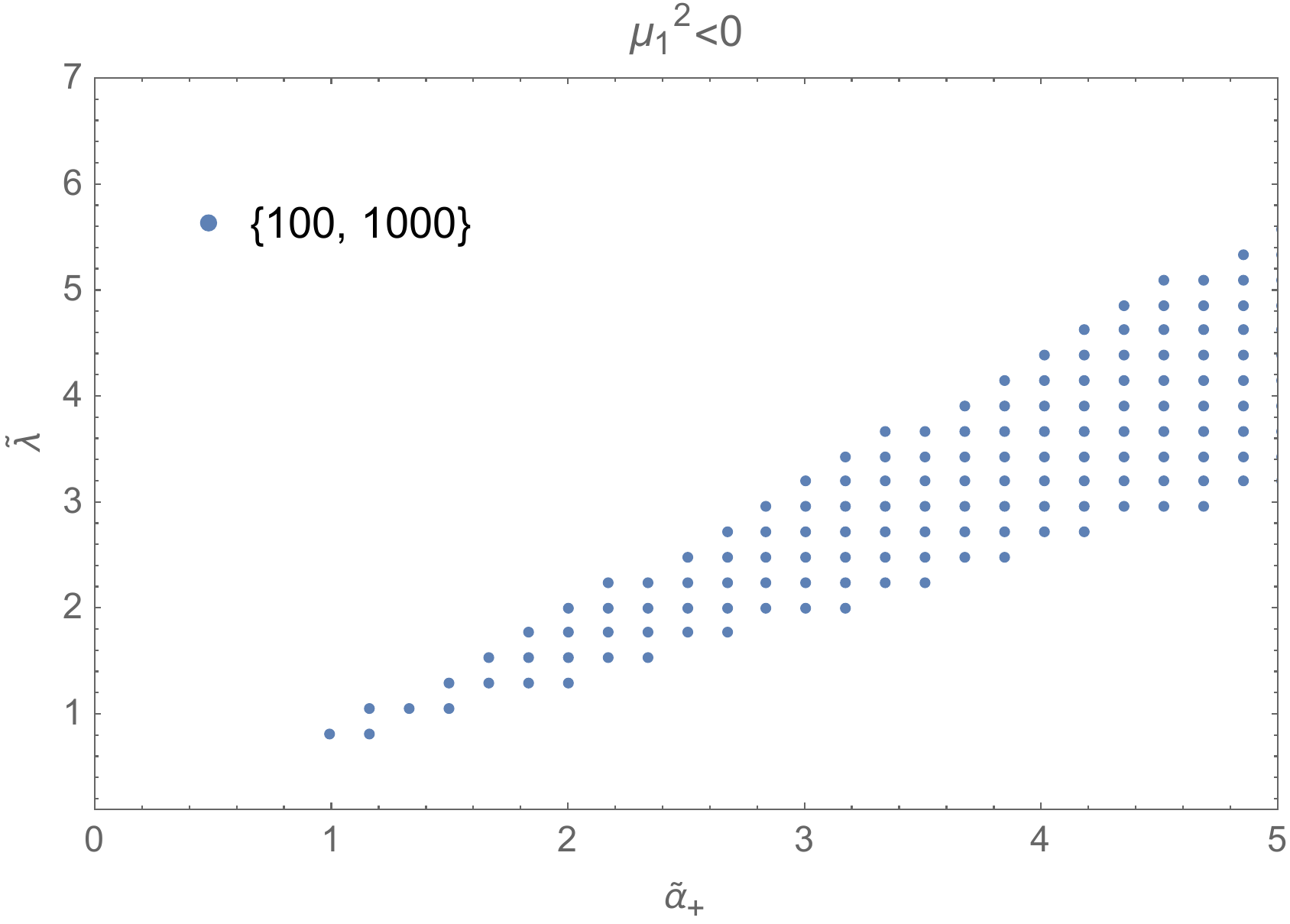}
        \caption{} 
        \label{alphapluslambda}
    \end{subfigure}
    \hspace{0.05cm}
    \begin{subfigure}[b]{0.49\textwidth}
        \centering
        \includegraphics[width=\textwidth]{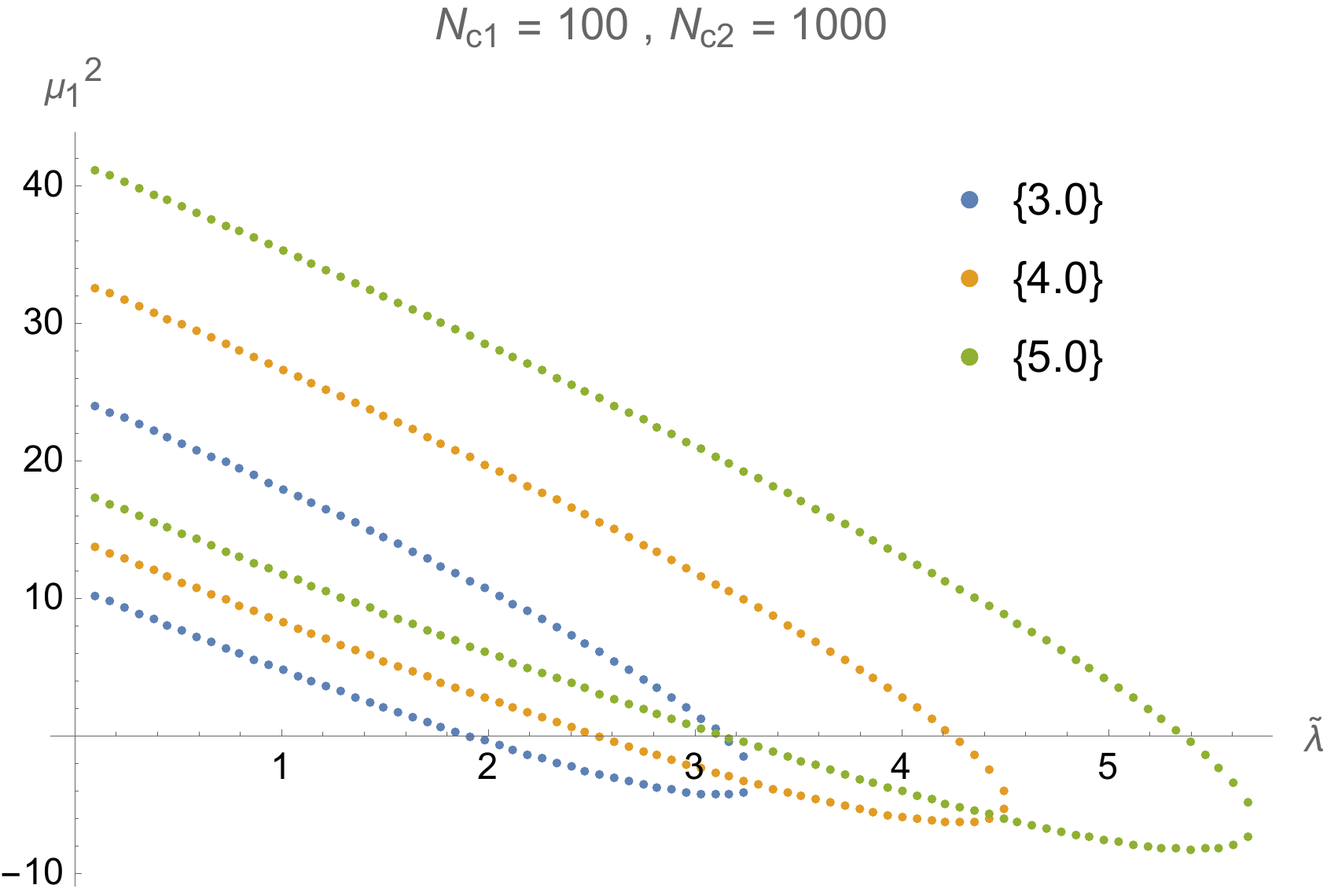}
        \caption{} 
        \label{lambdamu2}
    \end{subfigure}
    
    \caption{(a)  Region in the $\tilde\alpha_+$-$\tilde\lambda$ plane with negative thermal mass square for $(N_{c1},N_{c2})=(100,1000)$. This is in agreement with Fig.\ \ref{fig:regionplot}. (b) Mass square as a function of $\tilde\lambda$. Different colours represent different values of $\tilde\alpha_+$. For each $\tilde\alpha_+$ and $\tilde\lambda$ there are two solutions for the other parameters $\tilde\alpha_-$, $\tilde\lambda_{1,2}$.}
    \label{fig:regionplot2}
\end{figure}

The solutions are shown in Fig.\ \ref{fig:regionplot2}.

We still have to find explicitly $N_{f1,2}$ for which there is such a solution. For infinite $N_c$, this is always possible since $N_f/N_c$ is continuous, but in our case $N_{f1,2}$ are integers. As a benchmark point, we find:
\beq
(N_{c1},N_{c2},N_{f1},N_{f2},\tilde\lambda_1,
\tilde\lambda_2,\tilde\lambda)=(100,1000,534,5342,2.74077,11.0171,3.57732)\ ,
\eeq

\noi
for which $\tilde\alpha_+=63050/13251=4.75813$, $\tilde\alpha_-=50/13251=0.0037733$ and

\beq
(\mu_1^2,\mu_2^2)=(-2.80429,34.0568)\ .
\eeq

\begin{figure}[h!]
    \centering
    \begin{subfigure}[b]{0.49\textwidth}
        \centering
        \includegraphics[width=\textwidth]{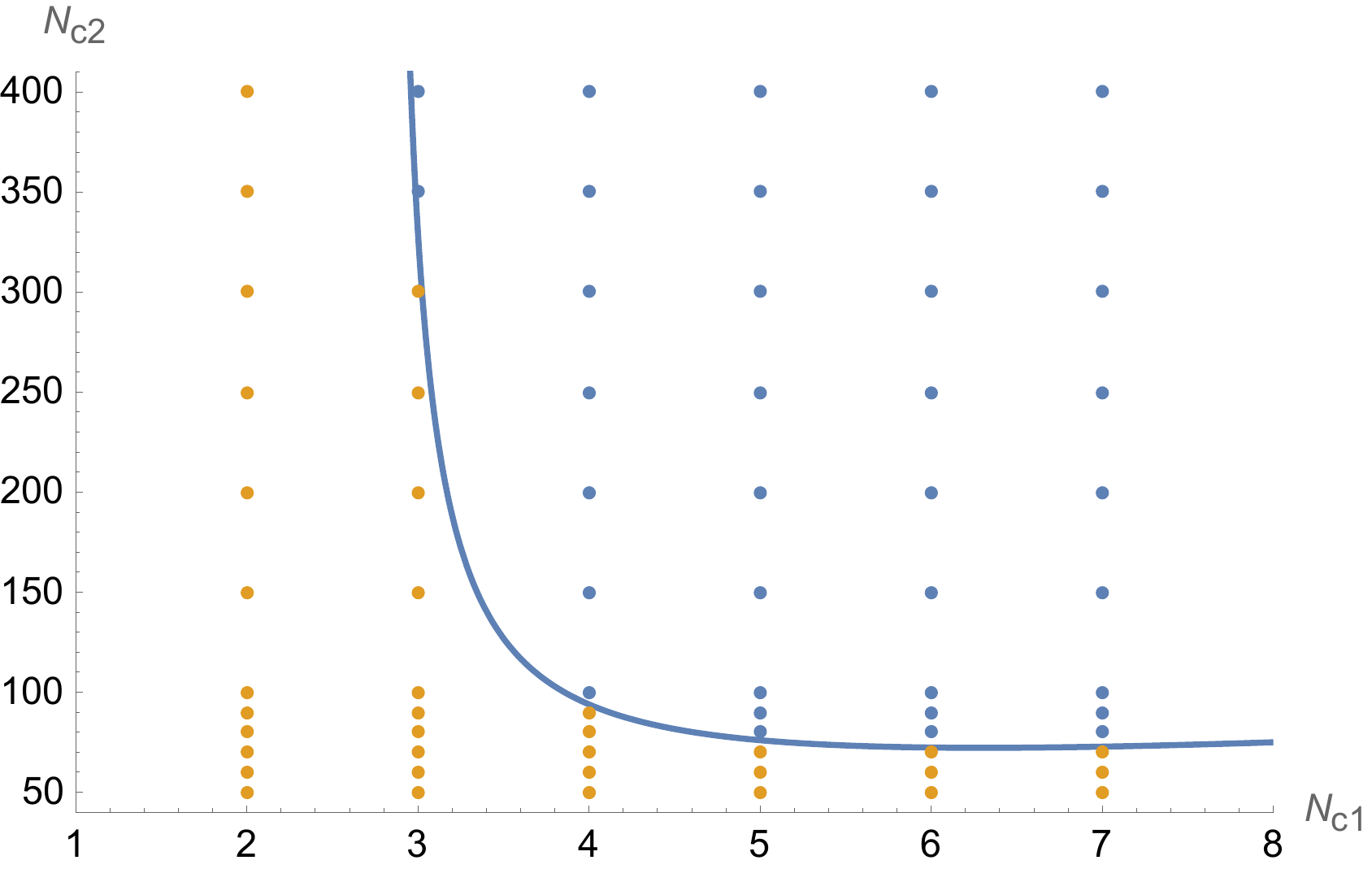}
        \caption{}
        \label{fig:regionplotNc}
    \end{subfigure}
    \hspace{0.05cm}
    \begin{subfigure}[b]{0.49\textwidth}
        \centering
        \includegraphics[width=\textwidth]{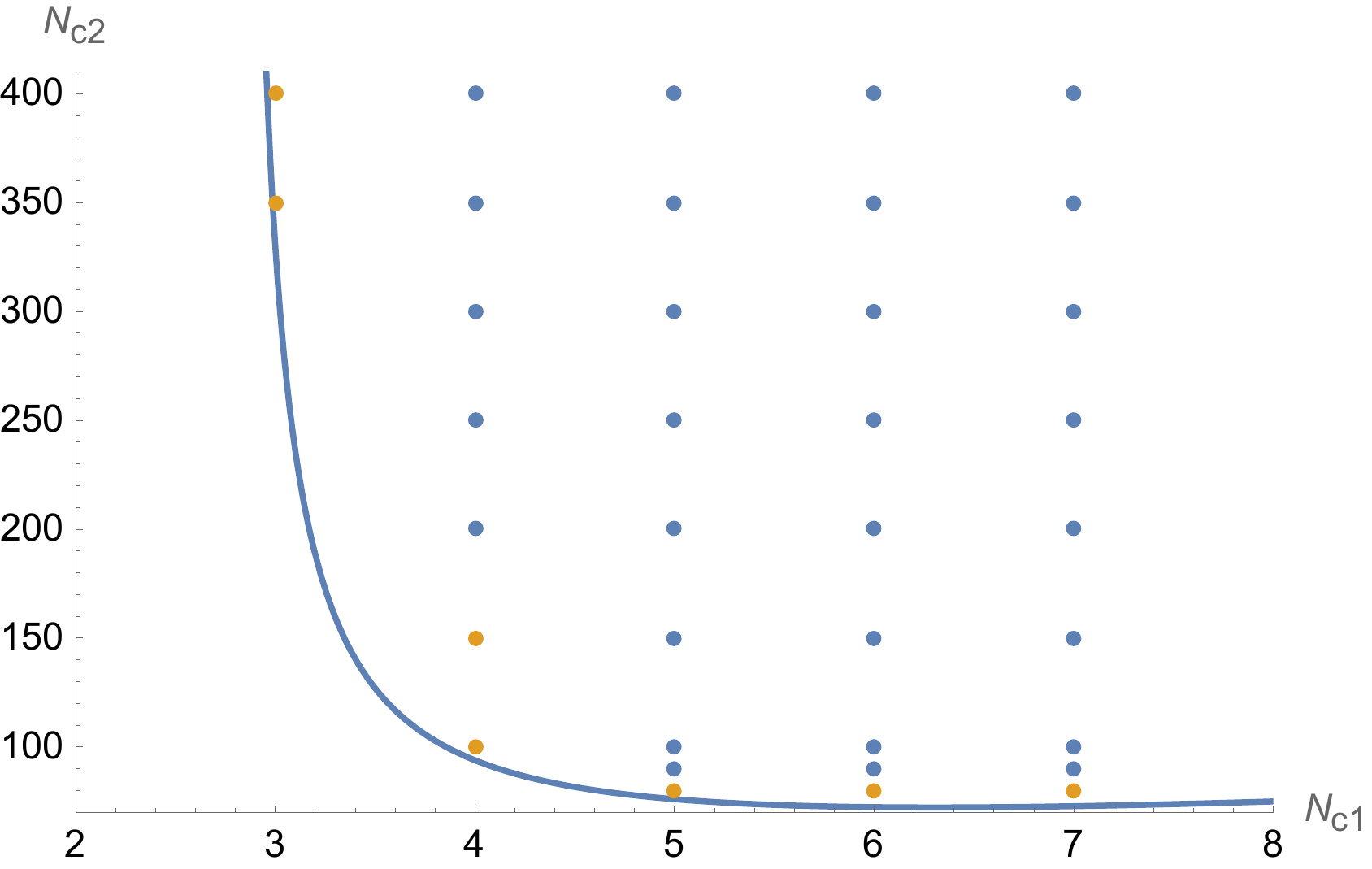}
        \caption{}
        \label{fig:regionplotNf}
    \end{subfigure}

    \caption{(a) Regions in the $N_{c1}-N_{c2}$ plane with (blue dots) and without (yellow dots) a solution with symmetry non-restoration. The boundary between them is well described by the constraint (\ref{N2}). Here $\tilde\alpha_{1,2}$ are arbitrary real numbers, not $1/(2b_{1,2})$. (b) Taking into account \eqref{eq:alpha12def} and varying integers $N_{f1,2}$, we plot the possible solutions above the blue continuous curve in panel (a). We see that some blue points, which represented allowed solutions before, become yellow, i.e. they are not solutions. This happens near the lower boundary.}
    \label{fig:regionplotNcNf}
\end{figure}

\noi
The resulting regions in the $N_{c1}$-$N_{c2}$ plane admitting solutions with symmetry non-restoration are displayed in Fig.~\ref{fig:regionplotNc}, where the boundary is well described by the analytical constraint \eqref{N2}. Once the additional restriction to integer flavour numbers is imposed, some of the allowed points are lost, as shown in Fig.~\ref{fig:regionplotNf}.

\section{Conclusions}
\label{sec:conclusions}

We studied whether symmetry non-restoration at arbitrarily high temperature can occur in an ultraviolet-complete quantum field theory in four spacetime dimensions with a finite field content. The key point is that the strict $T\to\infty$ limit is meaningful only if the microscopic theory remains valid at arbitrarily short distances. This naturally leads to the study of completely asymptotically free theories.

Our starting point was the large-$N$ analysis of the $\mathrm{SU}(N_{c1})\times \mathrm{SU}(N_{c2})$ model with two scalar fundamentals and a quartic cross-coupling. In the Veneziano limit, fixed-flow solutions were previously shown to exist for which one scalar acquires a negative thermal mass at high temperature, thereby realising symmetry non-restoration. In this work, we asked whether this mechanism survives when the number of degrees of freedom is kept finite.

We derived the finite-$N$ fixed-flow equations and the corresponding thermal masses, and then analysed the system in two complementary ways. First, we expanded around the large-$N$ solution and obtained the leading $1/N$ corrections analytically. This shows that the region of parameter space with negative thermal mass is not an artefact of the Veneziano limit. Second, we solved the finite-$N$ equations numerically and identified explicit benchmark points with integer numbers of colours and flavours. These benchmarks satisfy asymptotic freedom and boundedness of the scalar potential while still exhibiting a negative thermal mass for one scalar field.

Our results, therefore, provide an explicit perturbative realisation of infinite heat order in a four-dimensional ultraviolet-complete quantum field theory with finite field content. The essential ingredient is the product gauge-group structure, which gives enough parametric freedom for complete asymptotic freedom and symmetry non-restoration to coexist. This sharply distinguishes the model from simpler single-gauge-factor constructions, where the corresponding constraints are typically too restrictive.

A natural question concerns the role of higher-loop corrections. In an asymptotically free theory, all couplings vanish as $1/\log T$ at high temperature. The one-loop fixed-flow analysis therefore captures the leading behaviour, and $n$-loop corrections to both the beta functions and the thermal effective potential enter at relative order $\tilde\alpha^{n-1}\sim 1/(\log T)^{n-1}$, which is parametrically suppressed in the $T\to\infty$ limit. As is standard in quantum field theory, the perturbative series is expected to be asymptotic rather than convergent. However, this does not affect our conclusions: what matters is that, for sufficiently high temperature, the coupling is arbitrarily small and any finite truncation of the series provides a controlled approximation. In particular, the sign of the thermal mass squared is determined at one loop and is a continuous function of the couplings; higher-loop contributions cannot reverse it once the effective expansion parameter becomes small enough.

What about gravity? Gravitational corrections are generally expected to become important near the Planck scale and to modify the running of the couplings, typically strengthening the asymptotic freedom of gauge and Yukawa interactions, while the fate of scalar self-couplings remains dependent on the specific ultraviolet completion of gravity, as summarised in \cite{Antipin:2013bya}. If, however, the underlying particle theory is already completely asymptotically free below the Planck scale, it is natural to expect it to enjoy an enhanced degree of robustness against such Planckian effects. From this viewpoint, the theories studied here, when interpreted as dark sectors coupled only through gravity, are expected to preserve their infinite heat order up to, and plausibly slightly beyond, the Planck scale, until a complete theory of quantum gravity takes over.

The framework presented here also suggests several directions for future work. It would be valuable to classify more systematically the finite-$N$ theories that exhibit this behaviour, to assess whether analogous mechanisms can be realised in phenomenologically motivated settings, and to explore the possible cosmological implications of dark and visible sectors that become increasingly ordered at high temperature.

\subsubsection*{Acknowledgments}
BB acknowledges the financial support from the Slovenian Research Agency
(research core funding No.~P1-0035). The work of G.M., F.S. and S.W.  is partially supported by the Carlsberg Foundation, Semper Ardens grant CF22-0922.

 \begin{appendices}
 
\section{\label{eqs} The 1-loop RG equations}
\subsection{\label{SUNc12} $SU(N_{c1}) \times SU(N_{c2})$  with two scalar fundamentals}

The model under consideration has the gauge symmetry $SU(N_{c1}) \times SU(N_{c2})$, 
where each scalar field $\varphi_i$ transforms in the fundamental 
representation of its respective $SU(N_{ci})$ and is a singlet under the other. 
The most general potential is
\beq
V=\frac{\lambda_1}{2}\left(\vec{\varphi}_1^*\cdot\vec{\varphi}_1\right)^2+\frac{\lambda_2}{2}\left(\vec{\varphi}_2^*\cdot\vec{\varphi}_2\right)^2-
\lambda\left(\vec{\varphi}_1^*\cdot\vec{\varphi}_1\right)\left(\vec{\varphi}_2^*\cdot\vec{\varphi}_2\right)\;.\label{eq:pot}
\eeq
To derive the one-loop RGEs, we follow the approach established in \cite{Bajc:2021}, to which we refer for the details.\\
Again, to ensure that all the couplings are asymptotically free, i.e., that the model exhibits complete asymptotic freedom (CAF), we parametrise the couplings with a factor $1/t$ 
\bea
i=1,2&:&g_i^2=\frac{16\pi^2\tilde\alpha_i}{N_{ci}t}\quad,\quad \lambda_i=\frac{16\pi^2\tilde\lambda_i}{N_{ci}t}\quad,\quad\lambda=\frac{16\pi^2\tilde\lambda}{\sqrt{N_{c1}N_{c2}}t}\;.
\eea
This ensures that the couplings all vanish at the same rate in the UV, a behaviour known as fixed-flow \cite{Giudice:2015}. With this parametrisation, the RGEs reduce to a set of coupled polynomial equations, which for the model here discussed are the ones in Eq. \eqref{eq:finiteNRG}.
Real solutions of these algebraic equations correspond to models exhibiting CAF, as long as the gauge coupling is asymptotically free, i.e. $b_0>0$. For the model to be physically viable, it must also satisfy \eqref{eq:Vbounded} such that its potential is bounded from below. \\
 Plugging in Eq.'s \eqref{eq:finiteNRG} the coupling redefinitions \eqref{eq:couplings} and solving perturbatively order by order, we have that the LO couplings must satisfy the system
\begin{equation}\label{eq:LORGeq1}
    -\tilde\lambda^{(0)}_1=2\tilde\lambda^{(0)2}_1+2\tilde\lambda^{(0)2}-6\tilde\alpha^{(0)}_1\tilde\lambda^{(0)}_1+\frac{3}{2}\tilde\alpha_1^{(0)2}\;,\\
\end{equation}
\begin{equation}\label{eq:LORGeq2}
    -\tilde\lambda^{(0)}_2=2\tilde\lambda_2^{(0)2}+2\tilde\lambda^{(0)2}-6\tilde\alpha^{(0)}_2\tilde\lambda^{(0)}_2+\frac{3}{2}\tilde\alpha_2^{(0)2}\;,\\
\end{equation}
\begin{equation}\label{eq:LORGeq3}
    -\tilde\lambda^{(0)}=2\left(\tilde\lambda^{(0)}_1+\tilde\lambda^{(0)}_2\right)\tilde\lambda^{(0)}-3\left(\tilde\alpha^{(0)}_1+\tilde\alpha^{(0)}_2\right)\tilde\lambda^{(0)}\; ,
\end{equation}
which is the system solved for the infinite-$N$ case in \cite{Bajc:2021}. On the other hand, the NLO couplings will need to satisfy the system of equations
\begin{equation}\label{eq:NLORGeqp}
    \begin{split}
        &39\tilde\alpha_+^{(0)2}+(1-4\tilde\lambda_-^{(0)})^2-12\tilde\alpha_+^{(0)}(1-4\tilde\lambda_-^{(0)})+x\Big[1+39\tilde\alpha_+^{(0)2}+6\tilde\alpha_+^{(1)2}+16\tilde\lambda^{(0)}\tilde\lambda^{(1)}\\
        &-12\tilde\alpha_+^{(0)}\Big(1+2\tilde\alpha_+^{(1)}+4\tilde\lambda_-^{(0)}\Big)+8\tilde\lambda_-^{(0)}\Big(1-3\tilde\alpha_-^{(1)}+2\tilde\lambda_-^{(0)}+2\tilde\lambda_-^{(1)}\Big)\Big]=0 \ \ \ ,
    \end{split}
\end{equation}
\begin{equation}\label{eq:NLORGeqm}
    \begin{split}
        &39\tilde\alpha_+^{(0)2}+(1-4\tilde\lambda_-^{(0)})^2-12\tilde\alpha_+^{(0)}(1-4\tilde\lambda_-^{(0)})-x\Big[1+39\tilde\alpha_+^{(0)2}+6\tilde\alpha_-^{(1)2}\Big(4\tilde\alpha_+^{(0)}-1\Big)\\
        &-12\tilde\alpha_+^{(0)}\Big(1+4\tilde\lambda_-^{(0)}\Big)+8\tilde\lambda_-^{(0)}\Big(1+3\tilde\alpha_+^{(1)}+2\tilde\lambda_-^{(0)}-2\tilde\lambda_+^{(1)}\Big)\Big]=0 \ \ \ ,
    \end{split}
\end{equation}
\begin{equation}\label{eq:NLORGeq3}
    \begin{split}
        &-1+6\tilde\alpha_+^{(0)}-8\sqrt{x}\tilde\lambda^{(0)}+4\tilde\lambda_-^{(0)}+x\Big(-1+6\tilde\alpha_+^{(0)}-12\tilde\alpha_+^{(1)}-4\tilde\lambda_-^{(0)}+8\tilde\lambda_+^{(1)}\Big)=0 \ \ \ .
    \end{split}
\end{equation}

\noindent
The thermal effective potential is given by
\beq
\Delta V_T=
\frac{T^2}{48}\sum_{i=1}^2\sum_{a=1}^{N_{ci}}\left(4\frac{\partial^2V}{\partial\varphi_i^a\partial\varphi_{ia}^*}
+6g_i^2\frac{N_{ci}^2-1}{N_{ci}}\varphi_i^a\varphi_{ia}^*\right)\;,
\eeq
which for the potential \eqref{eq:pot} yields
\begin{align}
\Delta V_T=(4\pi)^2\frac{T^2}{24\log{T}}&\Bigg(\left(2\left({\tilde\lambda_1\left(1+\frac{1}{N_{c1}}\right)}-\sqrt{\frac{N_{c2}}{N_{c1}}}\tilde\lambda\right)+3\tilde\alpha_1 \left(1 - \frac{1}{N_{c1}^2} \right)\right)\left(\vec{\varphi}_1^*\cdot\vec{\varphi}_1\right)\nonumber\\
&+\left(2\left(\tilde\lambda_2\left(1+\frac{1}{N_{c2}}\right)-\sqrt{\frac{N_{c1}}{N_{c2}}}\tilde\lambda\right)+3\tilde\alpha_2 \left(1 - \frac{1}{N_{c2}^2} \right)\right)\left(\vec{\varphi}_2^*\cdot\vec{\varphi}_2\right)\Bigg)\;.
\end{align}
\end{appendices}
\printbibliography

@article{Giudice:2015,
    author = "Giudice, Gian F. and Isidori, Gino and Salvio, Alberto and Strumia, Alessandro",
    title = "{Softened Gravity and the Extension of the Standard Model up to Infinite Energy}",
    eprint = "1412.2769",
    archivePrefix = "arXiv",
    primaryClass = "hep-ph",
    reportNumber = "CERN-PH-TH-2014-247, IFT-UAM-CSIC-14-127",
    doi = "10.1007/JHEP02(2015)137",
    journal = "JHEP",
    volume = "02",
    pages = "137",
    year = "2015"
}

@article{Komargodski:2024zmt,
    author = "Komargodski, Zohar and Popov, Fedor K.",
    title = "{Temperature-Resistant Order in 2+1 Dimensions}",
    eprint = "2412.09459",
    archivePrefix = "arXiv",
    primaryClass = "hep-th",
    doi = "10.1103/yb7d-6tvc",
    journal = "Phys. Rev. Lett.",
    volume = "135",
    number = "9",
    pages = "091602",
    year = "2025"
}

@article{Han:2025eiw,
    author = "Han, Yiqiu and Huang, Xiaoyang and Komargodski, Zohar and Lucas, Andrew and Popov, Fedor K.",
    title = "{Entropic order}",
    eprint = "2503.22789",
    archivePrefix = "arXiv",
    primaryClass = "cond-mat.stat-mech",
    doi = "10.1038/s41467-025-66797-3",
    journal = "Nature Commun.",
    volume = "17",
    number = "1",
    pages = "87",
    year = "2026"
}

@article{Bajc:2021,
    author = "Bajc, Borut and Lugo, Adri{\'a}n and Sannino, Francesco",
    title = "{Asymptotically free and safe fate of symmetry nonrestoration}",
    eprint = "2012.08428",
    archivePrefix = "arXiv",
    primaryClass = "hep-th",
    doi = "10.1103/PhysRevD.103.096014",
    journal = "Phys. Rev. D",
    volume = "103",
    pages = "096014",
    year = "2021"
}

@inproceedings{Senjanovic:1998xc,
    author = "Senjanovic, Goran",
    title = "{Rochelle salt: A Prototype of particle physics}",
    booktitle = "{1st International Conference on Particle Physics and the Early Universe}",
    eprint = "hep-ph/9805361",
    archivePrefix = "arXiv",
    reportNumber = "IC-98-48",
    doi = "10.1142/9789814447263_0062",
    pages = "437--445",
    month = "9",
    year = "1998"
}

@inproceedings{Bajc:1999cn,
    author = "Bajc, Borut",
    title = "{High temperature symmetry nonrestoration}",
    booktitle = "{3rd International Conference on Particle Physics and the Early Universe}",
    eprint = "hep-ph/0002187",
    archivePrefix = "arXiv",
    reportNumber = "NYU-TH-00-02-01",
    doi = "10.1142/9789812792129_0039",
    pages = "247--253",
    year = "2000"
}

@article{Weinberg:1974hy,
    author = {S. Weinberg},
    title = "{Gauge and Global Symmetries at High Temperature}",
    journal = {Phys. Rev. D},
    volume = {9},
    pages = {3357},
    year = {1974},
    doi = {10.1103/PhysRevD.9.3357}
}

@article{Mohapatra:1979qt,
    author = {R. N. Mohapatra and G. Senjanovi\'c},
    title = "{Soft CP Violation at High Temperature}",
    journal = {Phys. Rev. Lett.},
    volume = {42},
    pages = {1651},
    year = {1979},
    doi = {10.1103/PhysRevLett.42.1651}
}

@article{Mohapatra:1979vr,
    author = {R. N. Mohapatra and G. Senjanovi\'c},
    title = "{Broken Symmetries at High Temperature}",
    journal = {Phys. Rev. D},
    volume = {20},
    pages = {3390},
    year = {1979},
    doi = {10.1103/PhysRevD.20.3390}
}

@article{Langacker:1980kd,
    author = {P. Langacker and S. Y. Pi},
    title = "{Magnetic Monopoles in Grand Unified Theories}",
    journal = {Phys. Rev. Lett.},
    volume = {45},
    pages = {1},
    year = {1980},
    doi = {10.1103/PhysRevLett.45.1}
}

@article{Salomonson:1984rh,
    author = {P. Salomonson and B. S. Skagerstam and A. Stern},
    title = "{On the Primordial Monopole Problem in Grand Unified Theories}",
    journal = {Phys. Lett. B},
    volume = {151},
    pages = {243-246},
    year = {1985},
    doi = {10.1016/0370-2693(85)90843-3}
}

@article{Dvali:1995cj,
    author = {G. R. Dvali and A. Melfo and G. Senjanovi\'c},
    title = "{Is There a Monopole Problem?}",
    journal = {Phys. Rev. Lett.},
    volume = {75},
    pages = {4559-4562},
    year = {1995},
    doi = {10.1103/PhysRevLett.75.4559},
    eprint = {hep-ph/9507230},
    archivePrefix = {arXiv},
    primaryClass = {hep-ph}
}

@article{Dvali:1995cc,
    author = {G. R. Dvali and G. Senjanovi\'c},
    title = "{Is There a Domain Wall Problem?}",
    journal = {Phys. Rev. Lett.},
    volume = {74},
    pages = {5178-5181},
    year = {1995},
    doi = {10.1103/PhysRevLett.74.5178},
    eprint = {hep-ph/9501387},
    archivePrefix = {arXiv},
    primaryClass = {hep-ph}
}

@article{Mohapatra:1979zc,
    author = {R. N. Mohapatra and G. Senjanovi\'c},
    title = "{Cosmological Baryon Production in a `superconducting' Early Universe}",
    journal = {Phys. Rev. D},
    volume = {21},
    pages = {3470},
    year = {1980},
    doi = {10.1103/PhysRevD.21.3470}
}

@article{Kuzmin:1981bc,
    author = {V. A. Kuzmin and M. E. Shaposhnikov and I. I. Tkachev},
    title = "{Gauge Hierarchies and Unusual Symmetry Behavior at High Temperatures}",
    journal = {Phys. Lett. B},
    volume = {105},
    pages = {159-162},
    year = {1981},
    doi = {10.1016/0370-2693(81)91011-X}
}

@article{Kuzmin:1981ip,
    author = {V. A. Kuzmin and M. E. Shaposhnikov and I. I. Tkachev},
    title = "{Matter - Antimatter Domains in the Universe: a Solution of the Vacuum Walls Problem}",
    journal = {Phys. Lett. B},
    volume = {105},
    pages = {167-170},
    year = {1981},
    doi = {10.1016/0370-2693(81)91013-3}
}

@article{Kuzmin:1982hy,
    author = {V. A. Kuzmin and M. E. Shaposhnikov and I. I. Tkachev},
    title = "{Baryon Generation and Unusual Symmetry Behavior at High Temperatures}",
    journal = {Nucl. Phys. B},
    volume = {196},
    pages = {29-44},
    year = {1982},
    doi = {10.1016/0550-3213(82)90300-5},
    note = {[erratum: Nucl. Phys. B \textbf{202} (1982), 543-544]}
}

@article{Dodelson:1989ii,
    author = {S. Dodelson and L. M. Widrow},
    title = "{Baryon Symmetric Baryogenesis}",
    journal = {Phys. Rev. Lett.},
    volume = {64},
    pages = {340-343},
    year = {1990},
    doi = {10.1103/PhysRevLett.64.340}
}

@article{Dodelson:1991iv,
    author = {S. Dodelson and B. R. Greene and L. M. Widrow},
    title = "{Baryogenesis, Dark Matter and the Width of the Z}",
    journal = {Nucl. Phys. B},
    volume = {372},
    pages = {467-493},
    year = {1992},
    doi = {10.1016/0550-3213(92)90328-9}
}

@article{Lee:1995fb,
    author = {J. w. Lee and I. g. Koh},
    title = "{Inflation and Inverse Symmetry Breaking}",
    journal = {Phys. Rev. D},
    volume = {54},
    pages = {7153-7157},
    year = {1996},
    doi = {10.1103/PhysRevD.54.7153},
    eprint = {hep-ph/9506415},
    archivePrefix = {arXiv},
    primaryClass = {hep-ph}
}

@article{Linde:1976kh,
    author = {A. D. Linde},
    title = "{High Density and High Temperature Symmetry Behavior in Gauge Theories}",
    journal = {Phys. Rev. D},
    volume = {14},
    pages = {3345},
    year = {1976},
    doi = {10.1103/PhysRevD.14.3345}
}

@article{Haber:1981ts,
    author = {H. E. Haber and H. A. Weldon},
    title = "{Finite Temperature Symmetry Breaking as Bose-Einstein Condensation}",
    journal = {Phys. Rev. D},
    volume = {25},
    pages = {502},
    year = {1982},
    doi = {10.1103/PhysRevD.25.502}
}

@article{Benson:1991nj,
    author = {K. M. Benson and J. Bernstein and S. Dodelson},
    title = "{Phase Structure and the Effective Potential at Fixed Charge}",
    journal = {Phys. Rev. D},
    volume = {44},
    pages = {2480-2497},
    year = {1991},
    doi = {10.1103/PhysRevD.44.2480}
}

@article{Riotto:1997tf,
    author = {A. Riotto and G. Senjanovi\'c},
    title = "{Supersymmetry and Broken Symmetries at High Temperature}",
    journal = {Phys. Rev. Lett.},
    volume = {79},
    pages = {349-352},
    year = {1997},
    doi = {10.1103/PhysRevLett.79.349},
    eprint = {hep-ph/9702319},
    archivePrefix = {arXiv},
    primaryClass = {hep-ph}
}

@article{Liu:1993am,
    author = {J. Liu and G. Segre},
    title = "{Baryon Asymmetry of the Universe and Large Lepton Asymmetries}",
    journal = {Phys. Lett. B},
    volume = {338},
    pages = {259-262},
    year = {1994},
    doi = {10.1016/0370-2693(94)91375-7}
}

@article{Bajc:1997ky,
    author = {B. Bajc and A. Riotto and G. Senjanovi\'c},
    title = "{Large Lepton Number of the Universe and the Fate of Topological Defects}",
    journal = {Phys. Rev. Lett.},
    volume = {81},
    pages = {1355-1358},
    year = {1998},
    doi = {10.1103/PhysRevLett.81.1355},
    eprint = {hep-ph/9710415},
    archivePrefix = {arXiv},
    primaryClass = {hep-ph}
}

@article{Bajc:1998rd,
    author = {B. Bajc and A. Riotto and G. Senjanovi\'c},
    title = "{R - Charge Kills Monopoles}",
    journal = {Mod. Phys. Lett. A},
    volume = {13},
    pages = {2955-2964},
    year = {1998},
    doi = {10.1142/S0217732398003132},
    eprint = {hep-ph/9803438},
    archivePrefix = {arXiv},
    primaryClass = {hep-ph}
}

@article{Bajc:1999he,
    author = {B. Bajc and G. Senjanovi\'c},
    title = "{Large Lepton Number and High Temperature Symmetry Breaking in Mssm}",
    journal = {Phys. Lett. B},
    volume = {472},
    pages = {373-381},
    year = {2000},
    doi = {10.1016/S0370-2693(99)01432-X},
    eprint = {hep-ph/9907552},
    archivePrefix = {arXiv},
    primaryClass = {hep-ph}
}

@article{Kang:1991xa,
    author = {H. S. Kang and G. Steigman},
    title = "{Cosmological Constraints on Neutrino Degeneracy}",
    journal = {Nucl. Phys. B},
    volume = {372},
    pages = {494-520},
    year = {1992},
    doi = {10.1016/0550-3213(92)90329-A}
}

@article{Kinney:1999pd,
    author = {W. H. Kinney and A. Riotto},
    title = "{Measuring the Cosmological Lepton Asymmetry Through the Cmb Anisotropy}",
    journal = {Phys. Rev. Lett.},
    volume = {83},
    pages = {3366-3369},
    year = {1999},
    doi = {10.1103/PhysRevLett.83.3366},
    eprint = {hep-ph/9903459},
    archivePrefix = {arXiv},
    primaryClass = {hep-ph}
}

@article{Lesgourgues:1999wu,
    author = {J. Lesgourgues and S. Pastor},
    title = "{Cosmological Implications of a Relic Neutrino Asymmetry}",
    journal = {Phys. Rev. D},
    volume = {60},
    pages = {103521},
    year = {1999},
    doi = {10.1103/PhysRevD.60.103521},
    eprint = {hep-ph/9904411},
    archivePrefix = {arXiv},
    primaryClass = {hep-ph}
}

@article{Barenboim:2016shh,
    author = {G. Barenboim and W. H. Kinney and W. I. Park},
    title = "{Resurrection of Large Lepton Number Asymmetries from Neutrino Flavor Oscillations}",
    journal = {Phys. Rev. D},
    volume = {95},
    number = {4},
    pages = {043506},
    year = {2017},
    doi = {10.1103/PhysRevD.95.043506},
    eprint = {1609.01584},
    archivePrefix = {arXiv},
    primaryClass = {hep-ph}
}

@article{Barenboim:2016lxv,
    author = {G. Barenboim and W. H. Kinney and W. I. Park},
    title = "{Flavor Versus Mass Eigenstates in Neutrino Asymmetries: Implications for Cosmology}",
    journal = {Eur. Phys. J. C},
    volume = {77},
    number = {9},
    pages = {590},
    year = {2017},
    doi = {10.1140/epjc/s10052-017-5147-4},
    eprint = {1609.03200},
    archivePrefix = {arXiv},
    primaryClass = {astro-ph.CO}
}

@article{Barenboim:2017dfq,
    author = {G. Barenboim and W. I. Park},
    title = "{A Full Picture of Large Lepton Number Asymmetries of the Universe}",
    journal = {JCAP},
    volume = {04},
    pages = {048},
    year = {2017},
    doi = {10.1088/1475-7516/2017/04/048},
    eprint = {1703.08258},
    archivePrefix = {arXiv},
    primaryClass = {hep-ph}
}

@article{Haber:1982nb,
    author = {H. E. Haber},
    title = "{Baryon Asymmetry and the Scale of Supersymmetry Breaking}",
    journal = {Phys. Rev. D},
    volume = {26},
    pages = {1317},
    year = {1982},
    doi = {10.1103/PhysRevD.26.1317}
}

@article{Mangano:1984dq,
    author = {M. L. Mangano},
    title = "{Global and Gauge Symmetries in Finite Temperature Supersymmetric Theories}",
    journal = {Phys. Lett. B},
    volume = {147},
    pages = {307-310},
    year = {1984},
    doi = {10.1016/0370-2693(84)90122-9}
}

@article{Bajc:1996kj,
    author = {B. Bajc and A. Melfo and G. Senjanovi\'c},
    title = "{On Supersymmetry at High Temperature}",
    journal = {Phys. Lett. B},
    volume = {387},
    pages = {796-800},
    year = {1996},
    doi = {10.1016/0370-2693(96)01113-6},
    eprint = {hep-ph/9607242},
    archivePrefix = {arXiv},
    primaryClass = {hep-ph}
}

@misc{Dvali:1998ct,
    author = "Dvali, G. R. and Krauss, Lawrence M.",
    title = "{High temperature symmetry breaking, SUSY flat directions, and the monopole problem}",
    eprint = "hep-ph/9811298",
    archivePrefix = "arXiv",
    reportNumber = "CWRU-P31-98, NYU-TH-98-10-05",
    month = "11",
    year = "1998"
}

@article{Bajc:1998jr,
    author = {B. Bajc and G. Senjanovi\'c},
    title = "{High Temperature Symmetry Breaking via Flat Directions}",
    journal = {Phys. Rev. D},
    volume = {61},
    pages = {103506},
    year = {2000},
    doi = {10.1103/PhysRevD.61.103506},
    eprint = {hep-ph/9811321},
    archivePrefix = {arXiv},
    primaryClass = {hep-ph}
}

@article{Bimonte:1995sc,
    author = {G. Bimonte and G. Lozano},
    title = "{On Symmetry Nonrestoration at High Temperature}",
    journal = {Phys. Lett. B},
    volume = {366},
    pages = {248-252},
    year = {1996},
    doi = {10.1016/0370-2693(95)01395-4},
    eprint = {hep-th/9507079},
    archivePrefix = {arXiv},
    primaryClass = {hep-th}
}

@article{Orloff:1996yn,
    author = {J. Orloff},
    title = "{The UV Price for Symmetry Nonrestoration}",
    journal = {Phys. Lett. B},
    volume = {403},
    pages = {309-315},
    year = {1997},
    doi = {10.1016/S0370-2693(97)00552-2},
    eprint = {hep-ph/9611398},
    archivePrefix = {arXiv},
    primaryClass = {hep-ph}
}

@article{Roos:1995vm,
    author = {T. G. Roos},
    title = "{Wilson Renormalization Group Study of Inverse Symmetry Breaking}",
    journal = {Phys. Rev. D},
    volume = {54},
    pages = {2944-2959},
    year = {1996},
    doi = {10.1103/PhysRevD.54.2944},
    eprint = {hep-th/9511073},
    archivePrefix = {arXiv},
    primaryClass = {hep-th}
}

@article{Pietroni:1996zj,
    author = {M. Pietroni and N. Rius and N. Tetradis},
    title = "{Inverse Symmetry Breaking and the Exact Renormalization Group}",
    journal = {Phys. Lett. B},
    volume = {397},
    pages = {119-125},
    year = {1997},
    doi = {10.1016/S0370-2693(97)00150-0},
    eprint = {hep-ph/9612205},
    archivePrefix = {arXiv},
    primaryClass = {hep-ph}
}

@article{AmelinoCamelia:1996sd,
    author = {G. Amelino-Camelia},
    title = "{Rayleigh-Ritz Variational Approximation and Symmetry Nonrestoration}",
    journal = {Phys. Lett. B},
    volume = {388},
    pages = {776-782},
    year = {1996},
    doi = {10.1016/S0370-2693(96)01199-9},
    eprint = {hep-ph/9610262},
    archivePrefix = {arXiv},
    primaryClass = {hep-ph}
}

@article{Pinto:1999pg,
    author = {M. B. Pinto and R. O. Ramos},
    title = "{A Nonperturbative Study of Inverse Symmetry Breaking at High Temperatures}",
    journal = {Phys. Rev. D},
    volume = {61},
    pages = {125016},
    year = {2000},
    doi = {10.1103/PhysRevD.61.125016},
    eprint = {hep-ph/9912273},
    archivePrefix = {arXiv},
    primaryClass = {hep-ph}
}

@article{Bimonte:1995xs,
    author = {G. Bimonte and G. Lozano},
    title = "{Can Symmetry Nonrestoration Solve the Monopole Problem?}",
    journal = {Nucl. Phys. B},
    volume = {460},
    pages = {155-166},
    year = {1996},
    doi = {10.1016/0550-3213(95)00626-5},
    eprint = {hep-th/9509060},
    archivePrefix = {arXiv},
    primaryClass = {hep-th}
}

@article{Fujimoto:1984hr,
    author = {Y. Fujimoto and S. Sakakibara},
    title = "{On Symmetry Nonrestoration at High Temperature}",
    journal = {Phys. Lett. B},
    volume = {151},
    pages = {260-262},
    year = {1985},
    doi = {10.1016/0370-2693(85)90847-0}
}

@article{Klimenko:1988ng,
    author = {K. G. Klimenko},
    title = "{1/N expansion in the O(N) x O(N) scalar theory and the problem of symmetry restoration at high temperature}",
    journal = {Theor. Math. Phys.},
    volume = {80},
    pages = {929-935},
    year = {1989},
    doi = {10.1007/BF01016185}
}

@article{Grabowski:1990qc,
    author = {M. P. Grabowski},
    title = "{The Effective potential and symmetry breaking in the O(N) x O(N) model}",
    journal = {Z. Phys. C},
    volume = {48},
    pages = {505-510},
    year = {1990},
    doi = {10.1007/BF01572032}
}

@article{Gavela_1998,
   title={Fading of symmetry nonrestoration at finite temperature},
   volume={59},
   ISSN={1089-4918},
   url={http://dx.doi.org/10.1103/PhysRevD.59.025008},
   DOI={10.1103/physrevd.59.025008},
   number={2},
   journal={Physical Review D},
   publisher={American Physical Society (APS)},
   author={Gavela, M. B. and Pène, O. and Rius, N. and Vargas-Castrillón, S.},
   year={1998},
   month=dec }

@article{Bimonte:1998he,
    author = {G. Bimonte and D. Iniguez and A. Tarancon and C. L. Ullod},
    title = "{A Monte Carlo Study of Inverse Symmetry Breaking}",
    journal = {Phys. Rev. Lett.},
    volume = {81},
    pages = {750-753},
    year = {1998},
    doi = {10.1103/PhysRevLett.81.750},
    eprint = {hep-lat/9802022},
    archivePrefix = {arXiv},
    primaryClass = {hep-lat}
}

@article{Hong:2000rk,
    author = {S. I. Hong and J. B. Kogut},
    title = "{Symmetry Nonrestoration in a Gross-Neveu Model with Random Chemical Potential}",
    journal = {Phys. Rev. D},
    volume = {63},
    pages = {085014},
    year = {2001},
    doi = {10.1103/PhysRevD.63.085014},
    eprint = {hep-th/0007216},
    archivePrefix = {arXiv},
    primaryClass = {hep-th}
}

@article{Chai:2020zgq,
    author = {N. Chai and S. Chaudhuri and C. Choi and Z. Komargodski and E. Rabinovici and M. Smolkin},
    title = "{Thermal Order in Conformal Theories}",
    journal = {Phys. Rev. D},
    volume = {102},
    number = {6},
    pages = {065014},
    year = {2020},
    doi = {10.1103/PhysRevD.102.065014},
    eprint = {2005.03676},
    archivePrefix = {arXiv},
    primaryClass = {hep-th}
}

@article{Buchel:2020thm,
    author = "Buchel, Alex",
    title = "{Thermal order in holographic CFTs and no-hair theorem violation in black branes}",
    eprint = "2005.07833",
    archivePrefix = "arXiv",
    primaryClass = "hep-th",
    doi = "10.1016/j.nuclphysb.2021.115425",
    journal = "Nucl. Phys. B",
    volume = "967",
    pages = "115425",
    year = "2021"
}

@article{Buchel:2020jfs,
    author = "Buchel, Alex",
    title = "{Fate of the conformal order}",
    eprint = "2011.11509",
    archivePrefix = "arXiv",
    primaryClass = "hep-th",
    doi = "10.1103/PhysRevD.103.026008",
    journal = "Phys. Rev. D",
    volume = "103",
    number = "2",
    pages = "026008",
    year = "2021"
}

@article{Wilson:1971bg,
    author = {K. G. Wilson},
    title = "{Renormalization group and critical phenomena. 1. Renormalization group and the Kadanoff scaling picture}",
    journal = {Phys. Rev. B},
    volume = {4},
    pages = {3174-3183},
    year = {1971},
    doi = {10.1103/PhysRevB.4.3174}
}

@article{Wilson:1971dh,
    author = {K. G. Wilson},
    title = "{Renormalization group and critical phenomena. 2. Phase space cell analysis of critical behavior}",
    journal = {Phys. Rev. B},
    volume = {4},
    pages = {3184-3205},
    year = {1971},
    doi = {10.1103/PhysRevB.4.3184}
}

@article{Gross:1973id,
    author = "Gross, David J. and Wilczek, Frank",
    editor = "Taylor, J. C.",
    title = "{Ultraviolet Behavior of Nonabelian Gauge Theories}",
    doi = "10.1103/PhysRevLett.30.1343",
    journal = "Phys. Rev. Lett.",
    volume = "30",
    pages = "1343--1346",
    year = "1973"
}

@article{Politzer:1973fx,
    author = "Politzer, H. David",
    editor = "Taylor, J. C.",
    title = "{Reliable Perturbative Results for Strong Interactions?}",
    doi = "10.1103/PhysRevLett.30.1346",
    journal = "Phys. Rev. Lett.",
    volume = "30",
    pages = "1346--1349",
    year = "1973"
}

@inproceedings{Weinberg:1976xy,
    author = "Weinberg, Steven",
    title = "{Critical Phenomena for Field Theorists}",
    booktitle = "{14th International School of Subnuclear Physics: Understanding the Fundamental Constitutents of Matter}",
    reportNumber = "HUTP-76-160",
    doi = "10.1007/978-1-4684-0931-4_1",
    month = "8",
    year = "1976"
}

@article{Chai:2020onq,
    author = "Chai, Noam and Chaudhuri, Soumyadeep and Choi, Changha and Komargodski, Zohar and Rabinovici, Eliezer and Smolkin, Michael",
    title = "{Symmetry Breaking at All Temperatures}",
    doi = "10.1103/PhysRevLett.125.131603",
    journal = "Phys. Rev. Lett.",
    volume = "125",
    number = "13",
    pages = "131603",
    year = "2020"
}

@article{Carlstr_m_2015,
   title={Spontaneous breakdown of time-reversal symmetry induced by thermal fluctuations},
   volume={91},
   ISSN={1550-235X},
   url={http://dx.doi.org/10.1103/PhysRevB.91.140504},
   DOI={10.1103/physrevb.91.140504},
   number={14},
   journal={Physical Review B},
   publisher={American Physical Society (APS)},
   author={Carlström, Johan and Babaev, Egor},
   year={2015},
   month=apr }

@article{Antipin:2013bya,
    author = "Antipin, Oleg and Krog, Jens and Mojaza, Matin and Sannino, Francesco",
    title = "{Stable E$\chi$ tensions with(out) gravity}",
    eprint = "1311.1092",
    archivePrefix = "arXiv",
    primaryClass = "hep-ph",
    reportNumber = "CP3-ORIGINS-2013-44, DIAS-2013-44",
    doi = "10.1016/j.nuclphysb.2014.06.023",
    journal = "Nucl. Phys. B",
    volume = "886",
    pages = "125--134",
    year = "2014"
}

@article{Buchel:2021ead,
    author = "Buchel, Alex",
    title = "{Compactified holographic conformal order}",
    eprint = "2107.05086",
    archivePrefix = "arXiv",
    primaryClass = "hep-th",
    doi = "10.1016/j.nuclphysb.2021.115605",
    journal = "Nucl. Phys. B",
    volume = "973",
    pages = "115605",
    year = "2021"
}

@article{Buchel:2021yay,
    author = "Buchel, Alex",
    title = "{A bestiary of black holes on the conifold with fluxes}",
    eprint = "2103.15188",
    archivePrefix = "arXiv",
    primaryClass = "hep-th",
    doi = "10.1007/JHEP06(2021)102",
    journal = "JHEP",
    volume = "06",
    pages = "102",
    year = "2021"
}

@article{Buchel:2022zxl,
    author = "Buchel, Alex",
    title = "{The quest for a conifold conformal order}",
    eprint = "2205.00612",
    archivePrefix = "arXiv",
    primaryClass = "hep-th",
    doi = "10.1007/JHEP08(2022)080",
    journal = "JHEP",
    volume = "08",
    pages = "080",
    year = "2022"
}

@article{Buchel:2023zpe,
    author = "Buchel, Alex",
    title = "{Holographic conformal order with higher derivatives}",
    eprint = "2312.15764",
    archivePrefix = "arXiv",
    primaryClass = "hep-th",
    doi = "10.1016/j.nuclphysb.2024.116578",
    journal = "Nucl. Phys. B",
    volume = "1004",
    pages = "116578",
    year = "2024"
}

@article{Chai:2021tpt,
    author = "Chai, Noam and Dymarsky, Anatoly and Goykhman, Mikhail and Sinha, Ritam and Smolkin, Michael",
    title = "{A model of persistent breaking of continuous symmetry}",
    eprint = "2111.02474",
    archivePrefix = "arXiv",
    primaryClass = "hep-th",
    doi = "10.21468/SciPostPhys.12.6.181",
    journal = "SciPost Phys.",
    volume = "12",
    number = "6",
    pages = "181",
    year = "2022"
}

@article{Chai:2021djc,
    author = "Chai, Noam and Dymarsky, Anatoly and Smolkin, Michael",
    title = "{Model of Persistent Breaking of Discrete Symmetry}",
    eprint = "2106.09723",
    archivePrefix = "arXiv",
    primaryClass = "hep-th",
    doi = "10.1103/PhysRevLett.128.011601",
    journal = "Phys. Rev. Lett.",
    volume = "128",
    number = "1",
    pages = "011601",
    year = "2022"
}

@article{Chai:2020hnu,
    author = "Chai, Noam and Rabinovici, Eliezer and Sinha, Ritam and Smolkin, Michael",
    title = "{The bi-conical vector model at $1/N$}",
    eprint = "2011.06003",
    archivePrefix = "arXiv",
    primaryClass = "hep-th",
    doi = "10.1007/JHEP05(2021)192",
    journal = "JHEP",
    volume = "05",
    pages = "192",
    year = "2021"
}

@article{Nakayama:2021fgy,
    author = "Nakayama, Yu",
    title = "{On the Trace Anomaly of the Chaudhuri{\textendash}Choi{\textendash}Rabinovici Model}",
    eprint = "2101.02861",
    archivePrefix = "arXiv",
    primaryClass = "hep-th",
    reportNumber = "RUP-21-1",
    doi = "10.3390/sym13020276",
    journal = "Symmetry",
    volume = "13",
    number = "2",
    pages = "276",
    year = "2021"
}

@article{Chaudhuri:2021dsq,
    author = "Chaudhuri, Soumyadeep and Rabinovici, Eliezer",
    title = "{Symmetry breaking at high temperatures in large N gauge theories}",
    eprint = "2106.11323",
    archivePrefix = "arXiv",
    primaryClass = "hep-th",
    doi = "10.1007/JHEP08(2021)148",
    journal = "JHEP",
    volume = "08",
    pages = "148",
    year = "2021"
}

\end{document}